\documentclass[12pt]{article}
\usepackage{cite}
\usepackage[subrefformat=parens]{subcaption}
\usepackage[T1]{fontenc}
\usepackage[utf8]{inputenc}
\usepackage[english]{babel}

\usepackage{graphicx}
\usepackage[dvipsnames]{xcolor}
\usepackage{float}

\usepackage{amsmath}
\usepackage{mathrsfs}
\usepackage{amsfonts}
\usepackage{amssymb}
\usepackage{amstext}
\usepackage{slashed}
\usepackage{braket}
\usepackage{caption}
\usepackage{subcaption}
\usepackage{layouts}
\usepackage[normalem]{ulem}
\usepackage{physics}

\usepackage[bookmarks, breaklinks, colorlinks,urlcolor=black, citecolor=red, linkcolor=blue]{hyperref}

\usepackage{array}

\def\dsp{\displaystyle}
\def\bea{\begin{align}}
\def\eea{\end{align}} 
\def\be{\begin{equation}}
\def\ee{\end{equation}} 
\def\nn{\nonumber}

\def\Im{\text{Im}}
\def\tev{\ensuremath{\mathrm{Te\kern -0.1em V}}}
\def\gev{\ensuremath{\mathrm{Ge\kern -0.1em V}}}
\def\mev{\ensuremath{\mathrm{Me\kern -0.1em V}}}
\def\lB{\ensuremath{\lambda_{B_q}}}
\def\lBs{\ensuremath{\lambda_{B_s}}}

\begin{document}

\begin{flushright}
\end{flushright}

\vskip 2cm

	\begin{center}
		
		{\Large\bf Probing the inverse moment of $B_s$-meson  distribution \\[1ex] amplitude via  $B_s \to \eta_s$ form factors} \\[8mm]
		{Rusa Mandal\,\footnote{Email: rusa.mandal@iitgn.ac.in},
        Praveen S Patil\,\footnote{Email: praveen.patil@iitgn.ac.in }
  and Ipsita Ray\,\footnote{Email: ipsita.r@iitgn.ac.in }}
\vskip 5pt
  {\small\em Indian Institute of Technology Gandhinagar, Department of Physics, \\ Gujarat 382355, India}

	\end{center}

\begin{abstract}

We investigate the inverse moment of the $B_s$-meson light-cone distribution amplitude (LCDA), denoted as $\lBs$ and defined within the heavy quark effective theory, through the calculation of $B_s \to \eta_s$ form factors. The presence of the $s$-quark inside the $B_s$-meson dictates a notable departure of approximately $20\%$ in the $\lBs$ value compared to the non-strange case $\lB$, as computed within the QCD sum rule approach, albeit with significant uncertainty. First, we compute the decay constant of the $\eta_s$-meson utilizing two-point sum rules while retaining finite $s$-quark mass contributions. Next, we constrain the parameter $\lBs$ by calculating $B_s \to \eta_s$ form factors within the light-cone sum rule approach, using $B_s$-meson LCDAs, and leveraging Lattice QCD estimates at zero momentum transfer from the HPQCD collaboration. Our findings yield $\lBs = 480\pm 92\, \mev$ when expressing the $B_s$-meson LCDAs in the Exponential model, consistent with previous QCD sum rule estimate yet exhibiting a 1.5-fold improvement in uncertainty. Furthermore, we compare the form factor predictions, based on the extracted $\lBs$ value, with earlier analyses for other channels such as $B_s \to D_s$ and $B_s \to K$.

\end{abstract}
\setcounter{footnote}{0} 

\newpage

\section{Introduction}

The difference between the light quark masses and strange quark mass leads to the violation of  $SU(3)_F$ symmetry. This has influenced various observations in the hadron data from the early days. The difference between the properties of $B_{u,d}$ and $B_s$-meson is one such example. For instance, the ratio $f_{B_s}/f_{B_{u,d}}$ computed from Lattice QCD exhibits a deviation of approximately 20\% from unity. Another crucial aspect of mesons lies in their internal non-perturbative structure, characterized by light-cone distribution amplitudes (LCDAs). These enter in the theory prediction of exclusive decays computed within certain approaches such as QCD factorization and heavy-quark expansion. The inverse moment of the leading twist distribution amplitude, $\lB$, is the key parameter to model the behaviour of all the LCDAs.

The estimate of $\lB$ is available within the framework of QCD sum rules which suffers from significantly large uncertainty~\cite{Braun:2003wx}. The first attempt to compute the inverse moment for $B_s$-meson LCDA namely, $\lBs$, can be found in Ref.~\cite{Khodjamirian:2020hob} where the effect of finite $s$-quark mass and difference between strange and non-strange quark condensate densities are systematically considered. In this study, the individual estimate was determined to be $\lBs= 438 \,\pm\, 150$ MeV, with the ratio $\lBs/\lB = 1.19 \,\pm\, 0.14$ demonstrating a 20\% departure from unity, indicative of $SU(3)_F$ violation\footnote{The finite $s$-quark mass effects in the general $B_s$-meson LCDA parametrization have recently been investigated in Ref.~\cite{Feldmann:2023aml} offering QCD sum rule compatible values for $\lBs$ within certain range of the model parameters.}. While in the future, accurate measurement of the photoleptonic mode $B \to \ell \nu \gamma$ will be able to provide better estimate of the parameter $\lB$ under the well-established factorization framework, the prospect for $\lBs$ is not promising in this regard as the corresponding channel for $B_s$, i.e., $B_s \to \ell \ell \gamma$ is prone to non-local contributions~\cite{Beneke:2020fot} that may contaminate the extraction of $\lBs$. Although one feasible approach involves utilizing the same inverse moment for non-strange and strange $B$-mesons in theoretical predictions while disregarding $SU(3)_F$ violation effects, this may lead to erroneous conclusions when comparing with the burgeoning dataset of $B_s$-meson decay modes in recent years. Consequently, there is a timely need to explore alternative methods to mitigate the uncertainty associated with the parameter $\lBs$. 

 In this work, we utilize inputs from Lattice QCD regarding the $\eta_s$-meson, a hypothetical $s\bar s$ state, whose mass is determined in terms of pion and kaon masses. This configuration is more cost-effective to simulate on the lattice and is employed to calibrate the $s$-quark mass in analyses involving strange mesons. The mass and decay constant values are reported as $m_{\eta_s}=0.6885(22)\,\gev$ and $f_{\eta_s}= 0.181 (55)$ GeV, respectively~\cite{Dowdall:2013rya}. We can regard this $s\bar s$ contribution as one component of the familiar $\eta-\eta^\prime$ states, where the other component consists of light quarks\footnote{For previous study on $B_s$-meson to $\eta-\eta^\prime$ form factor calculated using QCD sum rule method see Ref. \cite{Azizi:2010rx}.}. Notably, using data on the $\eta_s$ state enables us to extract pertinent information bypassing the complexities associated with octet-singlet mixing parameterization. 

Recently, the HPQCD collaboration~\cite{Parrott:2022rgu} has provided form factor estimates of several $B$ to light-meson states at the maximum recoil of the light-meson with greater accuracy. The results are then used in Ref.~\cite{Mandal:2023lhp} to provide indirect extraction of $\lB$ using light-cone sum rule (LCSR) framework where a notable two times improvement in the uncertainty is obtained. In similar spirit, here we compute the $B_s \to \eta_s$ form factors within LCSR approach and analyse the Lattice estimates to constrain the parameter $\lBs$. To the best of our knowledge, this represents the first indirect determination of the inverse moment of the leading twist $B_s$-meson LCDA.

The paper is organized as follows. We first compute the two-point sum rules for the $\eta_s$-meson decay constant in the QCD sum rule framework in Sec.~\ref{sec:2pt}. The prediction for decay constant with the inputs used are also provided there. In Sec.~\ref{sec:LCSR}, we first briefly outline the computation of $B_s \to P$ form factors, where $P$ is any pseudoscalar meson, using $B_s$-meson LCDAs. We present the result on $\lBs$ extraction and compare the form factor predictions for several decay modes in Sec.~\ref{sec:results}. Finally in Sec.~\ref{sec:summary} we summarize the outcomes. Several appendices contain the useful expressions e.g., LCDA parametrization in Appendix~\ref{app:BDA}, coefficients of the LCSRs for form factors in Appendix~\ref{app:C_coeff} and correlation matrices of the fitted coefficients in Appendix~\ref{app:coeff} obtained in this analysis  for the $B_s \to P$ form factors.

\section{Two-point sum rule for decay constant}
\label{sec:2pt}

In this section, we start by computing the two-point sum rule for the decay constant of the $\eta_s$-meson. Subsequently, we conduct a numerical analysis to provide an estimate for the decay constant. The ranges determined for certain input parameters, such as the effective threshold and Borel parameter, may be applied in the following section, where we calculate the form factors for the $B_s \to \eta_s$ transition.
The two-point correlation function of the axial-vector current $J_{\mu} =  \Bar{s} \gamma_{\mu}\gamma_{5} s$ reads
\begin{align}
    \Pi_{\mu \nu} (q^2) & = i \int d^4x e^{iqx} \langle 0|\mathcal{T}\{J_{\mu}(x)J^{\dagger}_{\nu}(0)\}|0 \rangle \nn \\ 
    & = (-g_{\mu \nu}q^2+ q_{\mu}q_{\nu})\Pi_T(q^2)+q_{\mu}q_{\nu}\Pi_L(q^2)\,,
    \label{eq:correl2pt}
\end{align}
where in the second line the correlation function is decomposed into two tensor structures whose coefficients are Lorentz invariant amplitudes $\Pi_T$ and $\Pi_L$. The partial conservation of axial current dictates that pseudoscalar mesons contribute to $\Pi_L$ amplitude while axial-vector meson contributions are encoded in $\Pi_T$. Hence as $\eta_s$ is a pseudoscalar meson, contribution from the pseudoscalar states can be isolated by projecting out the correlation function as 
\begin{align}
    q^2 \Pi_L(q^2) = \dfrac{q^{\mu}q^{\nu}}{q^2}\Pi_{\mu \nu}\equiv \hat\Pi(q^2) \,.
\end{align}
For the rest of our analysis, we only focus on the $\hat\Pi$ part of the correlation function.

In order to derive the hadronic spectrum of the correlation function \eqref{eq:correl2pt}, we presume the lowest lying state is the pseudoscalar $(J^P=0^-)$ i.e.  $\eta_s$-meson. 
The hadronic dispersion relation for the two-point function can be written as 
\begin{align}
    \hat\Pi_{\rm had}(q^2) = \int_0^{\infty} ds \dfrac{\rho_{\rm had}(s)}{s-q^2} \,,
    \label{eq:dispHad}
\end{align}
where $\rho_{\rm had}(s)$ denotes the  hadronic spectral density $\rho_{\rm had}(s) = \dfrac{1}{\pi}\Im \hat\Pi_{\rm had}(s)$. Defining the decay constant $f_{\eta_s}$ for $\eta_s$-meson as 
\begin{align}
    \langle 0|\Bar{s}\gamma_{\mu}\gamma_{5}s|\eta_s (k) \rangle = i k_\mu f_{\eta_s}\,,
    \label{eq:eta_decayconst}
\end{align}
we write
\begin{align}
\label{eq:rho_had}
	\rho_{\rm had}(s) = m^2_{\eta_s}f^2_{\eta_s}\delta (s - m^2_{\eta_s})+  \rho_{\rm cont}(s) \Theta(s-s_{th})\,,
\end{align}

where the $\rho_{\rm cont}$ denotes the higher excited and continuum states, and $s_{th}$ the lowest threshold of the states heavier than $\eta_s$-meson. We then substitute the spectral density into the dispersion relation \eqref{eq:dispHad} and perform a Borel Transformation to remove the subtraction terms\footnote{For our choice of the current in \eqref{eq:correl2pt}, one subtraction $\hat{\Pi}(0)$ is sufficient for the convergence of the integral in the dispersion relation \eqref{eq:dispHad}.} and obtain exponentially suppressed contribution from $\rho_{\rm cont}$.
\begin{align}
    \hat\Pi_{\rm had} = f^2_{\eta_s} m^2_{\eta_s} e^{-m^2_{\eta_s}/M^2}+ \int_{s_{th}}^{\infty} ds e^{-s/M^2} \rho_{\rm cont} (s)\,,
    \label{eq:hadbor}
\end{align}
where $M^2$ is the Borel parameter.

Next we proceed to compute the local operator product expansion (OPE) of the correlator \eqref{eq:correl2pt}  which can be decomposed in two parts.  The leading effect of this expansion gives the perturbative contribution and the higher order terms are given by a series of QCD vacuum condensates multiplied by the corresponding Wilson coefficients.

\begin{align}
    \hat\Pi_{\rm OPE}(q^2) = \hat\Pi_{\rm pert}(q^2) + \hat\Pi_{\rm cond}(q^2) \,.
\end{align}

\begin{figure} [t]
\centering
\begin{subfigure}[b]{.25\linewidth}
\centering
\includegraphics[width=\linewidth]{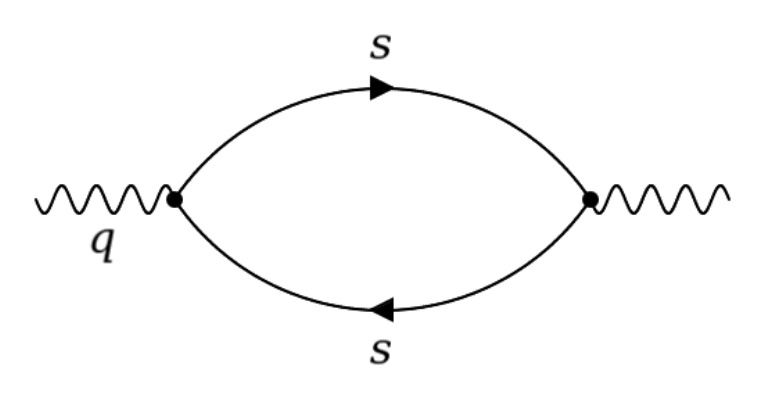}
\caption{}\label{fig:LO}
\end{subfigure} \hspace{1 cm}
\begin{subfigure}[b]{.25\linewidth}
\centering
\includegraphics[width=\linewidth]{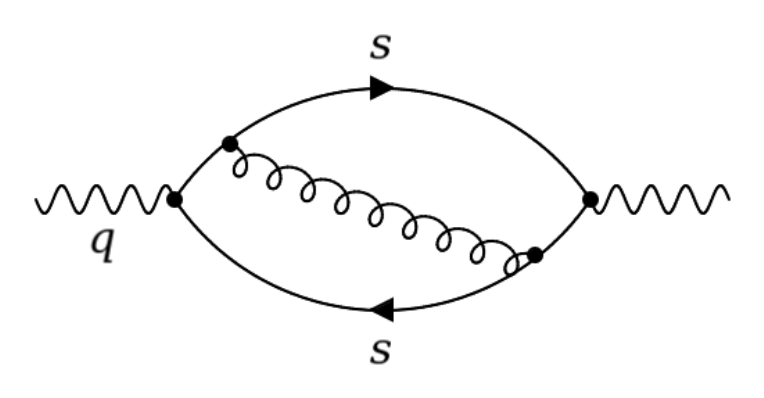}
\caption{}\label{fig:NLO1}
\end{subfigure} \hspace{1 cm}
\begin{subfigure}[b]{.25\linewidth}
\centering
\includegraphics[width=\linewidth]{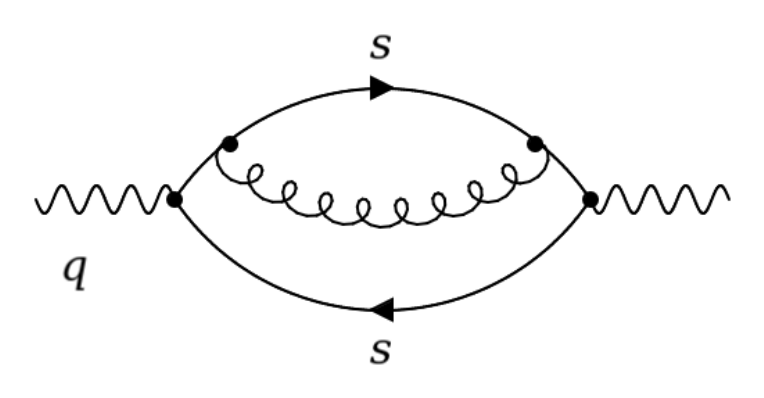}
\caption{}\label{fig:NLO2}
\end{subfigure}
\caption{\small Diagrams representing the perturbative part of the correlation function: (a) quark loop at leading order, (b) -- (c) the gluon radiative corrections at NLO ($\mathcal{O}(\alpha_s)$).}
\end{figure}

The spectral density of the correlator $\hat\Pi_{\rm pert}(q^2)$ is defined as $\rho_{\rm pert}(s) = \dfrac{1}{\pi}\Im \hat\Pi_{\rm pert}(s)$, and at leading order in $\alpha_s$, Fig. \ref{fig:LO} is
\begin{align}
   \rho_{\rm pert}^{\rm LO}(s) = \dfrac{3m_s^2}{2 \pi^2}\dfrac{\sqrt{s(s-4m_s^2)}}{s} \,.
\end{align}
The next-to-leading order (NLO) contribution to the perturbative spectral density $\rho_{\rm pert}^{\rm NLO}$  has been performed in Ref.~\cite{Generalis:1990id} for both vector and axial vector quark currents by retaining two distinct masses for the quarks. We use the expression with the suitable substitution for same quark masses. The corresponding diagrams are shown in Figs. \ref{fig:NLO1} and \ref{fig:NLO2}.

\begin{figure} [t]
\centering
\begin{subfigure}[b]{.25\linewidth}
\centering
\includegraphics[width=\linewidth]{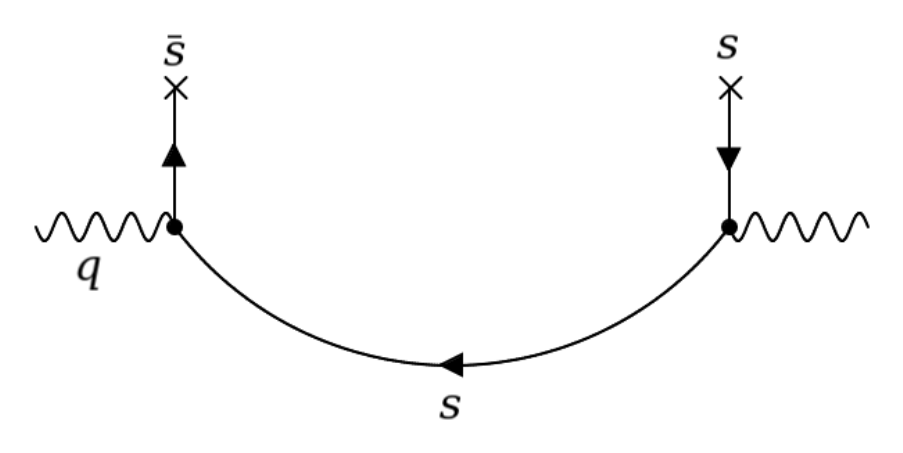}
\caption{}\label{fig:qq1}
\end{subfigure} \hspace{1 cm}
\begin{subfigure}[b]{.25\linewidth}
\centering
\includegraphics[width=\linewidth]{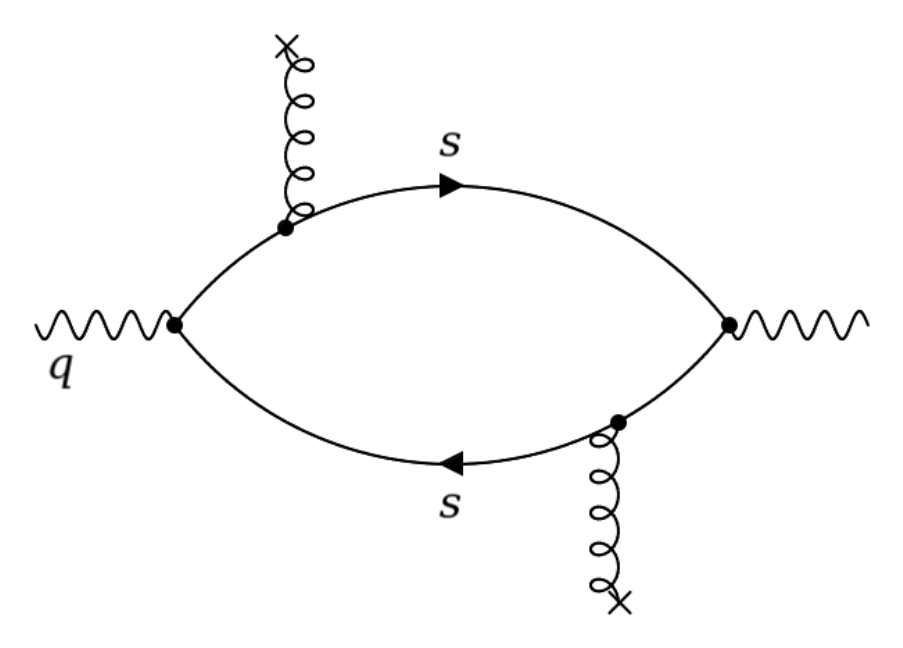}
\caption{}\label{fig:gg1}
\end{subfigure} \hspace{1 cm}
\begin{subfigure}[b]{.25\linewidth}
\centering
\includegraphics[width=\linewidth]{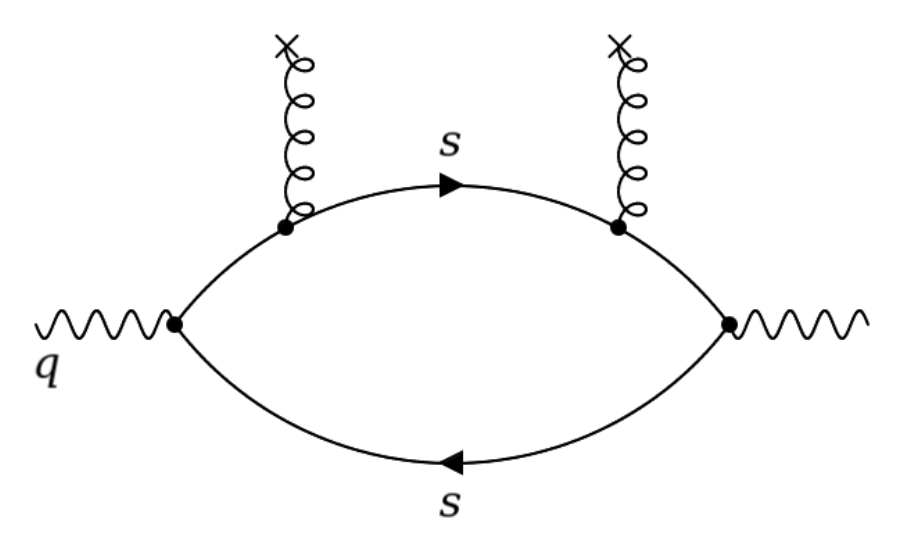}
\caption{}\label{fig:gg2}
\end{subfigure}
\begin{subfigure}[b]{.25\linewidth}
\centering
\includegraphics[width=\linewidth]{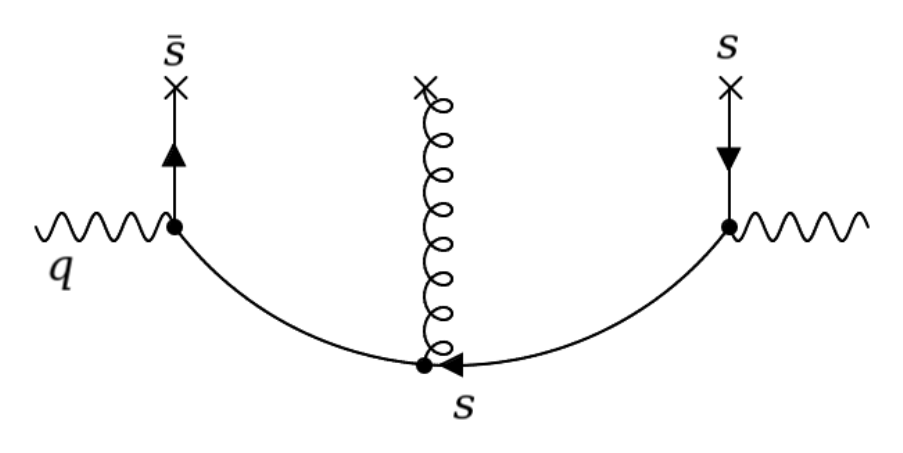}
\caption{}\label{fig:qg1}
\end{subfigure} \hspace{1 cm}
\begin{subfigure}[b]{.25\linewidth}
\centering
\includegraphics[width=\linewidth]{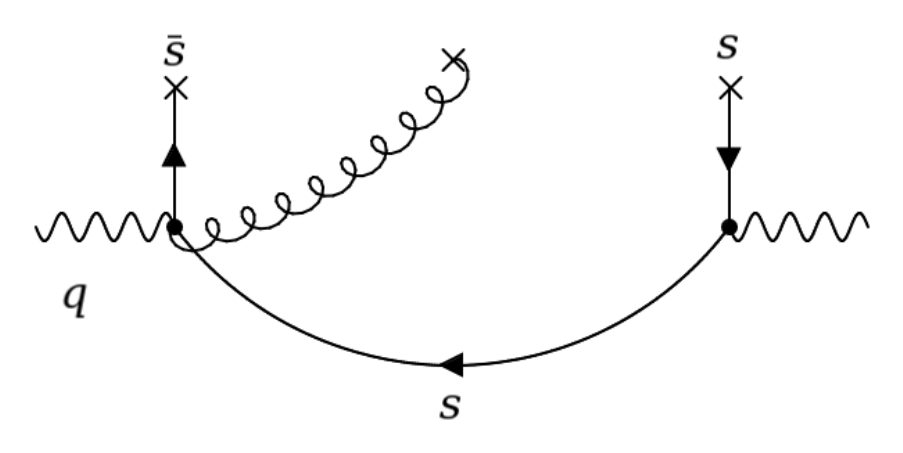}
\caption{}\label{fig:qg3}
\end{subfigure} 
\caption{\small Various contributions to the condensate terms: (a) $s\bar{s}$ quark condensate, (b) -- (c) gluon condensate and (d) -- (e) quark-gluon condensate contributions. Crossed lines denote vacuum fields.}
\end{figure}

The QCD vacuum condensates are organized in series with increasing mass dimension of the respective operators. We consider contributions up to $d=5$ condensates,
\begin{align}
    \hat\Pi_{\rm cond} = \hat\Pi^{\langle \Bar{q}q \rangle}+ \hat\Pi^{\langle GG \rangle}+ \hat\Pi^{ \langle \Bar{q}Gq \rangle}\,,
\end{align}
where $\hat\Pi^{\langle \Bar{q}q \rangle}$, $\hat\Pi^{\langle GG \rangle}$ and $\hat\Pi^{ \langle \Bar{q}Gq \rangle}$ are the quark, gluon and quark-gluon condensate contributions, respectively.
The lowest condensate contribution is at $d=3$ which is attributed to the case where one of the $s$-quark propagates perturbatively and the other one is absorbed and emitted by the $s$-quark condensate term, as shown in Fig. \ref{fig:qq1}. Expanding the quark fields as
\begin{equation}
 s(x)=s(0)+x^\mu \vec{D}_\mu s(0)+ \frac12 x^\mu x^\nu \vec{D}_\mu \vec{D}_\nu s(0) + \ldots \,,
\label{eq:qExpand}
\end{equation}
and taking the second term in the expansion, we obtain
\begin{align}
   \hat\Pi^{\langle \Bar{q}q \rangle} (q^2)= m_s   \left( \dfrac{3}{q^2 - m_s^2}+ \dfrac{q^2}{(q^2 - m^2_s)^2}\right) \langle \Bar{s}s \rangle\,.
\end{align}
Next arises the $d=4$ gluon condensate contributions in which we consider both the diagrams namely, one gluon emission from each of the two quark propagators, Fig. \ref{fig:gg1} and two gluon emission from either of the quarks, Fig. \ref{fig:gg2}. The resultant contribution to the correlator is \cite{Reinders:1984sr}
\begin{align}
\label{eq:GGcond_pi}
     \hat\Pi^{\langle GG \rangle} (q^2)= \left(-\dfrac{1}{12 m_s^2} - \dfrac{m_s^2}{4(-4m_s^2 +q^2)^2} -\dfrac{5}{24(-4m_s^2 +q^2)}\right) \langle GG \rangle\,.
\end{align}
Interestingly, apart from the contribution in Eq.~\eqref{eq:GGcond_pi}, we also generate imaginary contribution due to the presence of logarithms, which can be added to the OPE  using  dispersion relation (similar to the contribution to the perturbative part) of the following form.
\begin{align}
    \hat\Pi^{\langle GG \rangle} (q^2) = \langle GG \rangle \int_{4m_s^2}^{\infty} \dfrac{ds}{s-q^2}\rho^{\langle GG \rangle}(s)\, ,
\end{align}
where
\begin{align}
     \rho^{\langle GG \rangle}(s) = \dfrac{-30m_s^6 + 7m_s^4 s + m_s^2 s^2}{2s^4}\,.
\end{align}

We then take into account the effect of propagating quarks with quark-gluon condensate, where the vacuum expectation value of the quark-gluon-quark operator can be parametrized in terms of the parameter $m_0$ defined as
\begin{align}
    \langle 0 | \Bar{q}g_s G^a_{\mu \nu} t^a \sigma^{\mu \nu} q |0\rangle =\langle \Bar{q}Gq \rangle =  m_0^2 \langle \Bar{q}q \rangle\,.
    \label{eq:matele}
\end{align}
In this scenario, the contribution arises from two distinct effects; firstly, when the gluon emitted from the quark propagator ends in the vacuum, see Fig. \ref{fig:qg1}. Secondly, it stems from the term involving two covariant derivatives in the expansion of the quark field, Eq.~\eqref{eq:qExpand}. In the latter instance, the commutator of two covariant derivatives is proportional to the gluon field strength, which, upon combining with quark and anti-quark fields, forms the condensate. This can be interpreted as a gluon originating from the vacuum quark or anti-quark line, see Fig.~\ref{fig:qg3}. 
Summing these two effects, the resultant contribution is
\begin{align}
       \hat\Pi^{\langle \Bar{q}Gq \rangle}(q^2) = \dfrac{1}{2}\dfrac{m_s^3 }{(m_s^2- q^2)^3} \,m_0^2\langle\Bar{s}s\rangle\,.
\end{align}

After performing Borel Transformation in the variable $q^2$ of both the perturbative and condensate parts of the OPE, the correlation function is given by 
\begin{align}
\label{eq:Pi_OPE}
    \hat\Pi_{\rm OPE} = & \int_{4m_s^2}^{\infty}ds\, e^{-s/M^2}\left[ \rho^{\rm LO}_{\rm pert}(s)+ \dfrac{\alpha_s}{\pi} \rho^{\rm NLO}_{\rm pert}(s)+\rho^{\langle GG \rangle}(s)\langle GG \rangle \right]  \nn \\
    & -4m_se^{-m_s^2/M^2}\left(1- \dfrac{m^2_s}{4M^2} - \dfrac{m_s^2m_0^2}{16M^4}\right)\langle \bar ss \rangle + \left(\dfrac{5}{24} - \dfrac{m_s^2}{4M^2} \right)e^{-4m_s^2/M^2}\langle GG \rangle\,.
\end{align}
We are now at a stage to write down the QCD sum rules for the decay constant using  the quark-hadron duality approximation. In the asymptotic region of the spacelike momentum transfer, $q^2 \rightarrow - \infty$, the correlation function \eqref{eq:correl2pt} represents a short distance quark anti-quark fluctuation and can be treated in perturbative QCD. By making the use of Borel Transformation, which sets the momentum transferred variable $q^2$ to infinitely spacelike distances, the hadronic and perturbative parts of the correlation function can be made to coincide due to the absence of poles or cuts. We use semi-local quark-hadron duality assumption which dictates that the hadronic spectral density $\rho_{cont}(s)$, introduced in Eq.~\eqref{eq:rho_had} is equal to the perturbative part above some effective threshold $s_{th}$, as a consequence the integral over spectral density from $s_{th}$ to $\infty$ drops out while equating the 
OPE, Eq.~\eqref{eq:Pi_OPE} with the hadronic side, Eq.~\eqref{eq:hadbor}. The utilization of the Borel Transformation is crucial in this context, as it introduces an exponential suppression that reduces sensitivity to the assumption.
We then obtain  the sum rule for the decay constant for $\eta_s$-meson as
\begin{align}
   f^2_{\eta_s} m^2_{\eta_s} e^{-m^2_{\eta_s}/M^2} = & \int_{4m_s^2}^{s_{th}}ds\, e^{-s/M^2} \left[\rho^{\rm LO}_{\rm pert}(s)+ \dfrac{\alpha_s}{\pi} \rho^{\rm NLO}_{\rm pert}(s)+\rho^{\langle GG \rangle}(s)\langle GG \rangle \right] \nn \\
    - \,& 4m_se^{-m_s^2/M^2}\left(1- \dfrac{m^2_s}{4M^2} - \dfrac{m_s^2m_0^2}{16M^4}\right)\langle \bar ss \rangle + \left(\dfrac{5}{24} - \dfrac{m_s^2}{4M^2} \right)e^{-4m_s^2/M^2}\langle GG \rangle\,.
    \label{eq:feta_SR}
\end{align}

\begin{table}
\renewcommand{\arraystretch}{1.6}
\centering
\begin{tabular}{ |c|c|c|  }
 \hline
Parameter & Values &References\\
 \hline
 Normalization scale & $\mu = 1.5$ GeV & \cite{Gelhausen:2014jea, Khodjamirian_2021} \\
 \hline
 $s$-quark mass & $m_s(\mu=1.5\,\gev) =0.1016 \pm 0.0094$ GeV& \cite{Chetyrkin:2000yt,Workman:2022ynf} \\
 \hline
 Strong coupling & $\alpha_s(\mu=1.5\,\gev)=0.3494 \pm 0.0092 $  & \cite{Chetyrkin:2000yt, Workman:2022ynf}\\
 \hline
 Quark condensate   & $\langle \Bar{q}q \rangle =  (- 0.278 \pm 0.022 $ GeV)\textsuperscript{3}& \cite{FlavourLatticeAveragingGroupFLAG:2021npn} \\
 \hline
 $s$-quark condensate  &  $ \langle \Bar{s}s \rangle/ \langle \Bar{q}q \rangle$ =$(0.8\pm 0.3)$   & \cite{ioffe2003condensates}\\
 \hline
 Ratio $\langle \Bar{q}Gq\rangle/\langle\Bar{q}q\rangle$ &   $m_0^2 = 0.8 \pm 0.2 $
 GeV\textsuperscript{2}  & \cite{ioffe2003condensates}\\
 \hline
  Gluon condensates &  $\langle GG \rangle = 0.012 \pm 0.012 $  GeV\textsuperscript{4}& \cite{ioffe2003condensates}\\
  \hline
 $\eta_s$-meson mass & $m_{\eta_s}=0.6885 \pm 0.0022$ GeV & \cite{PhysRevD.103.094506}\\
 \hline
\end{tabular}
\caption{The input parameters used in two-point sum rule of the $\eta_s$-meson decay constant.}
\label{tab:inputs}
\end{table}

All the input parameters that are needed for the numerical evaluation are listed in Table~\ref{tab:inputs}.
The Borel parameter has to be chosen large enough such that the higher power correction in the OPE are sufficiently suppressed but small enough such that the contribution of continuum and excited states is subleading compared to the one of the lowest resonance. Also the sum rule should be independent of Borel parameter. These conditions are satisfied when the Borel parameter is in the range.
\begin{align}
    M^2 = [2,3] \,\text{GeV}\textsuperscript{2}\,.
\end{align}
The effective threshold $s_{th}$ is determined in the following way: we take the derivative of the sum rule with respect  to $-1/M^2$ and divide the result by the initial sum rule, which gives
\begin{align}
    m^2_{\eta_s} = \dsp\frac{\int_{4m_s^2}^{s_{th}}ds \,s\, e^{-s/M^2}\left[\rho^{\rm LO}_{\rm pert}(s)+ \dfrac{\alpha_s}{\pi} \rho^{\rm NLO}_{\rm pert}(s)+\rho^{\langle GG \rangle}(s)\langle GG \rangle\right]+ \dfrac{d \hat\Pi_{\rm cond}}{d\left(\frac{-1}{M^2}\right)}}{\int_{4m_s^2}^{s_{th}}ds\, e^{-s/M^2}\left[\rho^{\rm LO}_{\rm pert}(s)+ \dfrac{\alpha_s}{\pi}\rho^{\rm NLO}_{\rm pert}(s)+\rho^{\langle GG \rangle}(s)\langle GG \rangle \right]+ \hat\Pi_{\rm cond}}\,.
\end{align}
In the Borel window mentioned above, the effective threshold $s_{th}$ takes the value
\begin{align}
\label{eq:sth}
    s_{th} = (4.9 \pm 1.6)\, \text{GeV}^2\,,
\end{align}
and evaluating the sum rule \eqref{eq:feta_SR}, yields the decay constant
\begin{align}
    f_{\eta_s} = (187 \pm 37)\,\text{MeV}\,.
\end{align}

This value is consistent with the Lattice QCD estimate\,\footnote{as obtained by HPQCD collaboration $f_{\eta_s}^{\text{lat}}= (181 \pm 55)$\,MeV.} and the relation involving $K$-meson and pion decay constants, namely $\sqrt{2\,f_K^2 - f_\pi^2}$~\cite{Dowdall:2013rya}.

\section{Light-cone sum rules for form factors} 
\label{sec:LCSR}

In order to construct the LCSRs for $B_s \to \eta_s$ transition form factors, we start with the $B_s$-to-vacuum correlator
	\begin{align}
	\label{eq:correlator}
	\!\!\!\mathcal{F}^{\mu\nu}(q, k)
	= i\! \!\int\! \text{d}^4 x\, e^{i k\cdot x}\,
	\bra{0} \mathcal{T}\lbrace J_{int}^{\nu \dagger}(x), J_{weak}^{\mu}(0)\rbrace \ket{\bar{B}_s(q + k)},
	\end{align}
	of two quark currents. 
 Here, $J_{weak}^{\mu}(0) \equiv \bar{s}(0) \Gamma_w^\mu b(0)$ is the weak current with the momentum $q$ related to $B_s \to \eta_s$ transition. The interpolating current $J_{int}^{\nu} (x) \equiv \bar{s}(x) \gamma^\nu \gamma_5 s(x)$ is chosen to interpolate the $\eta_s$ with the momentum $k$. With our four-momenta assignment, the momentum of the $B_s$-meson state is $q+k$, which satisfies the on-shell condition, $p^2\equiv (q+k)^2=m_{B_s}^2$. 
 As a next step, we obtain the hadronic representation of the correlator ~\eqref{eq:correlator}
 via the dispersion relation. Taking the imaginary part with respect to the variable $k^2=s$ for $s>0$ and isolating the lowest lying $\eta_s$ state, it reads
 \begin{equation}
\begin{aligned}
    {\rm Im}_{k^2} {\cal F}_{\rm had}^{\mu\nu}(q,k)
    &=
    \pi
    \,\delta(s - m_{\eta_s}^2)
    \langle 0|J_{int}^{\nu \dagger}|\eta_s(k)\rangle
    \langle \eta_s(k) | J_{weak}^{\mu} | \bar{B}_s(q+k)\rangle
    +\dots
    \,,
\label{eq:ImcorrF}
\end{aligned}
\end{equation}
where the ellipsis denote contributions from other hadronic states with the same quantum
numbers.

 The first term in Eq.~\eqref{eq:ImcorrF} represents $\eta_s $ to vacuum transition and can be written in terms of the decay constant of the $\eta_s$-meson as defined in Eq.~\eqref{eq:eta_decayconst}.
 The second term denotes $B_s\to \eta_s$ hadronic matrix element for the weak current. Due to the conservation of parity, the axial-vector current transition matrix element vanishes. Introducing three independent non-vanishing form factors $f_+(q^2)$, $f_0(q^2)$ and $f_T(q^2)$ for the vector and the tensor weak currents i.e., $\Gamma^\mu_w\equiv \Gamma^\mu_V=\gamma^\mu$ and $\Gamma^\mu_w\equiv \Gamma^\mu_T=\sigma^{\mu \nu}q_\nu$, respectively, the hadronic matrix element can be parameterized. These form factors are functions of the momentum transfer $q^2 = (p - k)^2$, however, here we suppress the explicit $q^2$ dependence for simplicity.
	\begin{align}
\label{eq:BtoP:vector}
	\bra{\eta_s(k)} \bar{s} \gamma^\mu b \ket{B_s(p)}
	& = \left[(p + k)^\mu - \frac{m_{B_s}^2 - m_{\eta_s}^2}{q^2}q^\mu\right]\, f_+  
	+ \frac{m_{B_s}^2 - m_{\eta_s}^2}{q^2} q^\mu\, f_0,\\
	\label{eq:BtoP:tensor}
	\bra{\eta_s(k)} \bar{s} \sigma^{\mu\nu}\, q_\nu b \ket{B_s(p)}
	& = \frac{i f_T}{m_{B_s} + m_{\eta_s}}\, \left[q^2\, (p + k)^\mu  - (m_{B_s}^2 - m_{\eta_s}^2)\, q^\mu\right]\,.
	\end{align}
Substituting the definition of decay constant and the hadronic matrix elements Eqs.~\eqref{eq:BtoP:vector} and \eqref{eq:BtoP:tensor} for the vector and tensor currents respectively, in Eq.~\eqref{eq:ImcorrF} we get,
 \begin{align}
\label{eq:ImcorrFV}
    \frac1\pi{\rm Im}_{k^2} {\cal F}_{\rm had,V}^{\mu\nu}(q,k)
    &= i f_{\eta_s}
    \left[ \left\{ 2k^\mu k^\nu - \frac{m_{B_s}^2 - m_{\eta_s}^2-q^2}{q^2}q^\mu k^\nu \right\} f_+ \right. \nn \\
    & \hspace{1.7cm} \left. +\frac{m_{B_s}^2 - m_{\eta_s}^2}{q^2} q^\mu k^\nu \, f_0\right] \delta(s - m_{\eta_s}^2)
    +\dots
    \,,\\
\frac1\pi{\rm Im}_{k^2} {\cal F}_{\rm had,T}^{\mu\nu}(q,k)
    &=  
    \frac{-f_{\eta_s} f_T}{m_{B_s} + m_{\eta_s}}\, \left[2q^2 k^\mu k^\nu  - (m_{B_s}^2 - m_{\eta_s}^2-q^2)\, q^\mu k^\nu\right] \delta(s - m_{\eta_s}^2)
    +\dots
    \,
\label{eq:ImcorrFT}
\end{align}
Decomposing in different Lorentz structures e.g., $k^\mu k^\nu$ for $f_+$ and $f_T$, separately for two weak currents, and $q^\mu k^\nu$ for $f_\pm$ defined in Eq.~\eqref{eq:fpm}, we can extract the contributions by substituting Eqs.~\eqref{eq:ImcorrFV} and \eqref{eq:ImcorrFT}  for the form factor `$F$' in the dispersion relation
\begin{align}
    {\cal F}^{(F)}_{\rm had}(q^2,k^2)=\frac{1}{\pi} \int\limits_{0}^\infty \!ds 
    \,\frac{{\rm Im}_{k^2} {\cal F}^{(F)}_{\rm had}(s,q^2)}{s-k^2}\,.
\label{eq:dispF}
\end{align}

We now perform the OPE calculation of the same correlator \eqref{eq:correlator}.
For $k^2 \ll m_{s}^2$ and $q^2 \ll (m_b+m_{s})^2$, implying far below the hadronic thresholds in the channels of the interpolating and weak currents, the correlator \eqref{eq:correlator} can be calculated, expanding the
time-ordered product of currents near the light-cone $x^2 \simeq 0 $ as	%
	\begin{align}
	\label{eq:corrOPE}
	{\mathcal F}_{\rm OPE}^{\mu\nu}( q,k) = \!\int \!d^4 x\, e^{ik\cdot x}\!\!\int & \frac{d^4l}{(2\pi)^4}e^{-il\cdot x}
	\big[\gamma^\nu \gamma_5
	\frac{\slashed{l}+m_{s}}{m_{s}^2-l^2}
	\Gamma^\mu_w \big]_{\alpha\beta} \, \times \langle 0|\bar{s}^\alpha(x)b^\beta(0)  | \bar{B_s}(q+k)\rangle
	\,,  
	\end{align}
	where $\alpha,\beta$ are spinor indices. Here the virtual $s$-quark propagates between the two vertices of currents, $J_{int}^{\nu \dagger}(x)$ and $ J_{weak}^{\mu}(0)$, whereas a quark-antiquark pair emitted at a light-cone separation forms a $B_s$-meson state. We neglect the effect of soft-gluon emitted from the light-cone expansion of the $s$-quark propagator and absorbed, together with the quark-antiquark pair generating quark-antiquark-gluon (three-particle) components of the $B_s$-meson DAs. These effects are found to be negligible, in comparison to the two-particle contributions, for several $B$ to light-meson transition form factors~\cite{Gubernari:2018wyi}. The non-local $B$-to-vacuum matrix element $\left\langle 0\left|\bar{s}^\alpha(x) b^\beta(0)\right| \bar{B_s}(q+k)\right\rangle$ then can be approximated in terms of the two-particle $B_s$-meson LCDAs defined in the heavy quark effective theory, see Appendix~\ref{app:BDA} for the details. The inverse moment of the leading twist LCDA of the $B_s$-meson, defined as
\begin{equation}
\lambda_{B_s} = \Big{(}\int_0^{\infty} dw \frac{\phi_+(w)}{w} \Big{)}^{-1}\,,
\end{equation}
is the key parameter in parameterizing the LCDAs. The LCDAs of higher twist also depend on the parameters $\lambda_E$ and $\lambda_H$ which correspond to local matrix elements denoting quark-gluon three-body components in the $B_s$-meson wavefunction. These differ from the three-particle components of the LCDAs which occur in vacuum to meson matrix elements including non-local quark operators.
The renormalisation scale dependence of the LCDAs are neglected here and in the next section we determine the value of $\lBs$ at a reference scale $\mu=1\,\gev$.

 Evaluating the integral in Eq.~\eqref{eq:corrOPE}, after some redefinition of variables we can isolate the contribution for different form factors using the appropriate Lorentz structures as mentioned previously for the hadronic part, 
\begin{align}
\label{eq:OPEres}
    {\cal F}^{(F)}_{\rm OPE}(q^2, k^2)
    =
    f_{B_s} m_{B_s}\sum_{n=1}^{4}
    \int\limits_0^\infty d\sigma 
    \frac{
        I_n^{(F)}(\sigma,q^2)
    }{
        \left( k^2 -s \right)^n
    } \,,
\end{align}
 where 
 \begin{equation}
\sigma = \frac{\omega}{m_{B_s}},~~~~ s(\sigma,q^2) = \sigma m_{B_s}^2 + \frac{m_s^2 -\sigma q^2}{\bar{\sigma}}, ~~~~ \bar{\sigma} = 1-\sigma\,.
 \end{equation}
 The functions $I_n^{(F)}$ are combinations of two-particle  LCDAs $\phi_+, \phi_-, g_+$ and $g_-$ with twist two, three, four, and five, respectively, given as
 \begin{equation}
I_n^{(F,2p)}(\sigma,q^2) = \frac{1}{\bar{\sigma}^n} \sum_{\psi_{2 p}} C_n^{(F,\psi_{2p})}(\sigma,q^2) \psi_{2 p} (\sigma m_{B_s}), ~~~\psi_{2 p} = \phi_+, \bar{\phi}, g_+, \bar{g};
\end{equation}
with
\begin{equation}
		\bar{\phi}(w) = \int_0^w d \eta (\phi_+ (\eta) - \phi_-(\eta)) \,,~~
		\bar{g}(w) = \int_0^w d \eta (g_+(\eta) - g_-(\eta))\,.
\end{equation}
The expressions for the coefficients are collected in Appendix~\ref{app:C_coeff}.

 We are now at a stage to finally write the sum rules for all the form factors $F$ by equating the OPE correlation function \eqref{eq:OPEres} to the hadronic representation
in Eq.~\eqref{eq:dispF} with the assumption of semi-local quark-hadron duality. This can be written in a compact form \cite{Gubernari:2018wyi},
\begin{align} 
\label{eq:FF}
F= &\frac{f_{B_s} M_{B_s}}{K^{(F)}} \sum_{n=1}^{4}  
 \Big{\{}(-1)^{n}\int_{0}^{\sigma_0} d \sigma \;e^{(-s(\sigma,q^2)+m_{\eta_s}^2)/M^2} \frac{1}{(n-1)!(M^2)^{n-1}} I_n^{(F)}  \nonumber \\
    & - \Big{[}\frac{(-1)^{n-1}}{(n-1)!} e^{(-s(\sigma,q^2)+m_{\eta_s}^2)/M^2} \sum_{j=1}^{n-1} \frac{1}{(M^2)^{n-j-1}}\frac{1}{s'} (\frac{d}{d \sigma} \frac{1}{s'})^{j-1} I_n^{(F)} \Big{]}_{\sigma=\sigma_0} \Big{\}}\,,
\end{align}
where 
 \begin{equation}
s'(\sigma,q^2) = \frac{d s(\sigma,q^2)}{d \sigma}\,,
 \end{equation}
and $K^{(F)}$ are normalization factors, given in the appendix in Eq.~\eqref{eq:Kfac}.
In the above expression of form factors in Eq.~\eqref{eq:FF}, we performed the Borel transform in variable $k^2 \to M^2$. In the numerical section, the Borel window will be chosen such that the contributions from excited resonances and continuum states are exponentially suppressed and also the impact of higher-twist contributions, which varies as powers of 1/$M^2$ are suppressed. Here, $\sigma_0$ = $\sigma(s_0,q^2)$ where $s_0$ is the effective threshold which is expected to be close to the mass square of the first excited state.

\section{Numerical analysis and $\lBs$ extraction}
\label{sec:results}

In this section, at the first part, we perform the numerical analysis to estimate the form factors using the sum rules obtained in the last section and then in the second part, we make use of the Lattice QCD inputs to determine the parameter $\lBs$. As already discussed in Sec.~\ref{sec:LCSR}, the choice of the Borel parameter is an important concern in the calculation of the sum rules for the form factors. We have checked that those conditions are satisfied if we take $M^2$ = [2,\,3] $\text{GeV}^2$ for the $B_s \to \eta_s$ channel. To obtain the threshold parameter $s_0$, we follow the similar procedure described as of the two point sum rule for the decay constant. We take the derivative of the sum rule in Eq. \ref{eq:FF} for each form factor multiplied with 
$e^{-m_{\eta_s}^2/M^2}$ with respect to $-1/M^2$ and normalize it to the original sum rule. The result is equal to the mass square of the final state meson i.e.,
\begin{equation} \label{FF derivative}
m_{\eta_s}^2 = \frac{\frac{d}{d[-1/M^2]}[F e^{-m_{\eta_s}^2/M^2}]}{F e^{-m_{\eta_s}^2/M^2}}\,.
\end{equation}

In order to estimate the threshold parameter, we first quote the other inputs which enter in the form factor expressions via the $B_s$-meson LCDAs. The value for $\lambda_{B_s}$ has been taken from Ref.\cite{Khodjamirian:2020hob} and those of $\lambda_E^2$ and $\lambda_H^2$ are taken from Ref. \cite{Nishikawa:2011qk}. We form a multinormal distribution by varying these parameters, which follow a Gaussian distribution, within the ranges as given below.
\begin{equation} \label{eq:val}
	\lambda_{B_s} =  0.438 \pm 0.150~\rm GeV,~~~
	\lambda_E^2 =  0.03 \pm 0.02~\rm GeV^{2},~~~ 
	\lambda_H^2 =  0.06 \pm 0.03~\rm GeV^{2}.
\end{equation} 
  The Borel parameter $M^2$ follows a uniform distribution within $[2,\,3]\,\gev^2$. We generate 50 random points from the distribution and find the values of the threshold parameters satisfying Eq. \eqref{FF derivative} for each of the 50 combinations of the above mentioned parameters, thus leading to a distribution for the threshold parameter $s_0$ for the form factors $f_+$ and $f_T$ for four $q^2$ values in the range  $-15\,\text{GeV}^2$ to $0\, \text{GeV}^2$. We observe very mild $q^2$ dependence of the threshold parameters and the values are consistent within their corresponding $\pm1\,\sigma$ confidence interval (CI). From a conservative viewpoint, we take the average of the values obtained at the four different points of $q^2$ as our central value for the threshold parameter and the highest uncertainty as our uncertainty estimate as given below,
 \begin{equation}
	s_0 =\! 
	\left\{\hspace{-1mm}
	\begin{array}{lcl}
	\displaystyle 1.10 \pm 0.38\, \gev^2 &\hspace{-5mm} & (f_+,\,f_0), \label{eq:fplus}
	\\[2ex]
	1.23 \pm 0.31\,\gev^2  &\hspace{-5mm} & (f_T)\,  .
	\end{array}
	\right. 
\end{equation}

\begin{figure*}[h!!!!!]
		\small
		\centering
\includegraphics[width=0.32\textwidth]{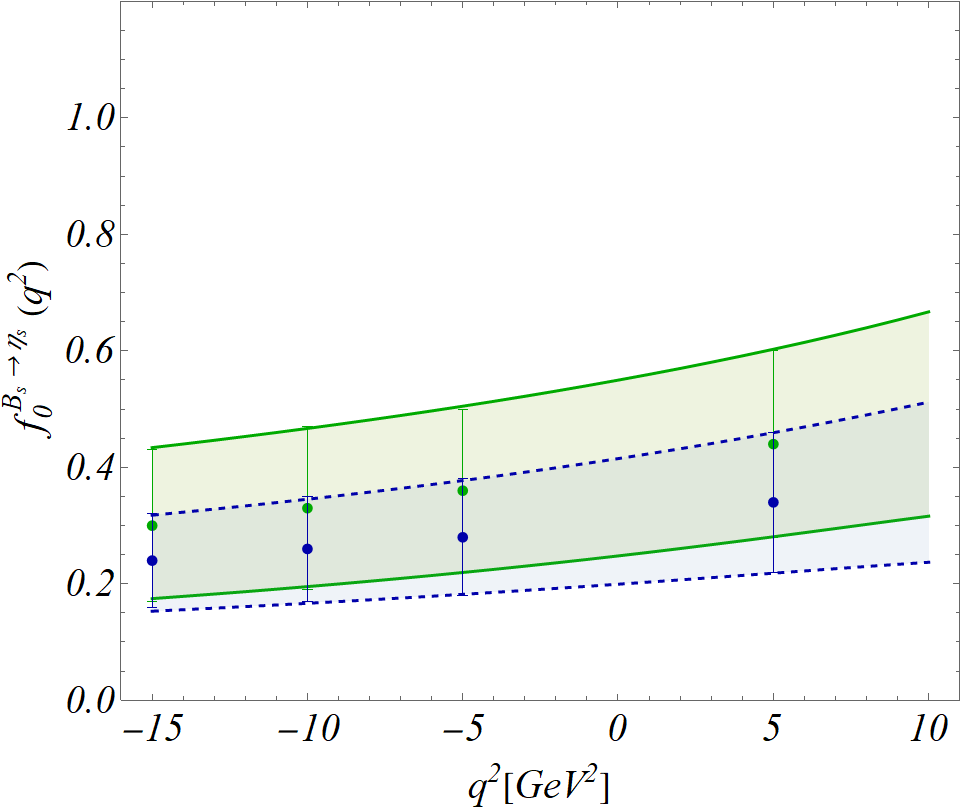}~~
\includegraphics[width=0.32\textwidth]{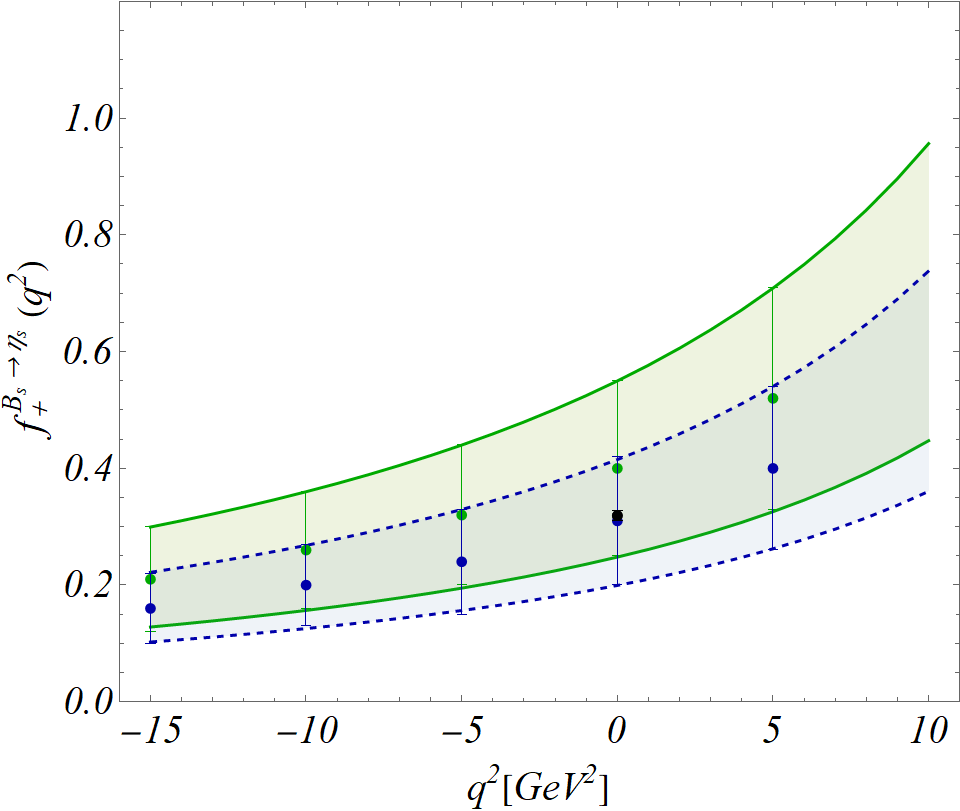}~~
\includegraphics[width=0.32\textwidth]{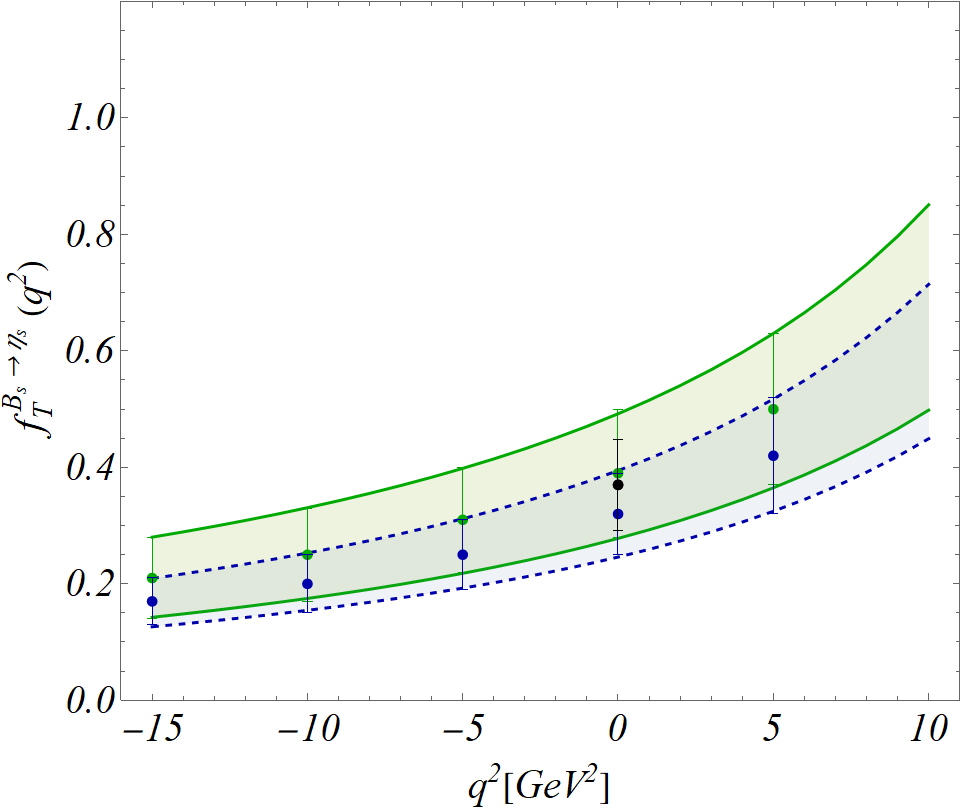}
\caption{\small The variation of $B_s \to \eta_s$ form factors $f_+,\,f_0$ and $f_T$ as a function of momentum transfer $q^2$. The green solid band with error bars represent the $\pm 1\sigma$ uncertainty envelope using QCD sum rule estimate of inverse moment parameter $\lBs$ in Eq.~\eqref{eq:val} whereas the blue dashed band and error bars denote the prediction using the extracted value of $\lBs$ in Eq.~\eqref{eq:lamval}. The data points for $f_{+,\,T}$ at $q^2$ = 0 $\text{GeV}^2$ from the HPQCD collaboration \cite{Parrott:2022rgu} are also shown in black.}
  \label{fig:FFBs2eta}
\end{figure*}

We are now at a stage to estimate the form factors for $B_s\to \eta_s$ transition using the sum rules obtained in Sec.~\ref{sec:LCSR}. We first create LCSR data points for the form factors $f_+$ and $f_T$ at $q^2 =\{ -15,\,-10,\,-5,\,0,\,5\}\,\text{GeV}^2$ and for $f_0$ at $q^2 = \{-15,\,-10,\,-5,\,5\}\,\text{GeV}^2$. For the extrapolation in the semileptonic region i.e., $4 m_{\ell}^2< q^2< (m_{B_s} - m_{\eta_s} )^2$, we perform a $q^2 \to z$ map by fitting these data points to a series expansion  in parameter $z$ 
\begin{align}
    \label{eq:zdef}
    z(q^2) = \frac{\sqrt{t_+-q^2} - \sqrt{t_+ - t_0^{\phantom{1}}}}{\sqrt{t_+-q^2} + \sqrt{t_+ - t_0^{\phantom{1}}}}
    \,,
\end{align}
where
\begin{align}
    & t_+ = (m_{B_s} + m_{\eta_s})^2 \,,&
    & t_0 = (m_{B_s} + m_{\eta_s})\left(\sqrt{m_{B_s}} - \sqrt{m_{\eta_s}}\right)^2 \,.&
\end{align}
The form factor then can be written as 
\begin{align}
     F^{(i)}_{\rm fit} (q^2)
    =
    \frac{1}{1 - \frac{q^2}{m_{(i)}^2}}
    \sum_{k=0}^2
    a_k^{(i)}
    \left[
        z(q^2)
        -
        z(0)
    \right]^k
    \,.
    \label{eq:zexpfit}
\end{align}
Here, $a_k^{(i)}$ are the form factor parameters\footnote{In this work, all the form factors are truncated at $k=2$.}. The resonance mass parameters, denoting the sub-thresholds with appropriate quantum numbers, used are
$m_{(0)}= 5.630\,\gev$ and $m_{(+,T)}= 5.412 \,\gev$ for $f_0$ and $f_{+,T}$, respectively \cite{Bharucha:2015bzk}.  The prediction of all three form factors as a function of $q^2$ is shown in Fig.~\ref{fig:FFBs2eta}. The green band denotes the $\pm 1\sigma$ uncertainty envelope.  The uncertainties of our calculation for $q^2 > 10$ $\text{GeV}^2$ further increase in the semileptonic region due to the extrapolation and hence we restrict the plot only up to that point.

Our next aim is to look for improvement in the uncertainty for the form factor predictions which is dominated by the highly uncertain parameter $\lBs$. In this regard, we perform the numerical analysis to extract the value of $\lambda_{B_s}$ using the form factor estimates at $q^2$ = 0 $\text{GeV}^2$ from the HPQCD collaboration \cite{Parrott:2022rgu}. We first define an
	optimized $\chi^2$- statistic as
	\begin{equation}
	\label{eq:chisq}
	\chi^2 = \sum_{i,j} (O_i^{\rm Lattice}- O_i^{\rm theo}).\, Cov^{-1}_{ij}. \,(O_j^{\rm Lattice}- O_j^{\rm theo}) + \chi^2_{\rm nuis}\,.
	\end{equation} 
	Here, $O_i^{\rm Lattice}$ corresponds to the Lattice inputs (central values) of the form factors $f_+$ and $f_T$ for the $B_s \to \eta_s$ transition at $q^2$ = 0\footnote{Note that at $q^2$ = 0, as $f_+(0)$ = $f_0(0)$, we do not include $f_0(0)$ in the $\chi^2$ statistic.}  and $Cov$ is the corresponding covariance matrix which includes the correlation in data. We have assigned $\chi^2$ corresponding to the nuisance parameters $\lambda_E^2$, $\lambda_H^2$, $s_0$ and $M^2$
	in $\chi^2_{\rm nuis}$ where the parameters $\lambda_E^2$, $\lambda_H^2$ and  $s_0$ follow Gaussian distributions, as discussed above with the corresponding 1$\sigma$ CIs. quoted in Eqs. \eqref{eq:val} and \eqref{eq:fplus}, whereas $M^2$ follows a uniform distribution. Note that while determining the effective threshold $s_0$ in Eq.~\eqref{eq:fplus}, the input value of $\lBs$ is used however we mention that the procedure of taking ratio (outlined in Eq.~\eqref{FF derivative}) reduces the dependency on the input parameters. Furthermore, by treating $s_0$ as a nuisance parameter and varying it in the $\chi^2$ fit, this effect is further minimized. Alternatively, one conventional approach is to set the value of $s_0$ the same as the effective threshold obtained in the analysis of the two-point sum rule for the decay constant, as seen in Eq.~\eqref{eq:sth}. However, due to its significant uncertainty, we opt not to use it here. 

 Using the statistical frequentist procedure as described in Ref. \cite{Mandal:2023lhp}, we obtain 

 \begin{equation}
 \lambda_{B_s} (1\,\gev) = (480^{+92}_{-83})\, \text{MeV}\,.
 \label{eq:lamval}
 \end{equation} 
Here, in this indirect extraction, a `reasonable' choice of the factorization scale $\mu =1\,\gev$ is assumed at which the values of other input parameters entering in the sum rule are considered.
 
 We now compare the predictions for the form factors for $B_s \to P (P = \eta_s, K, D_s)$ channels using our estimate of $\lambda_{B_s}$ with the earlier analyses. The LCSR data points are created using the analytic expression \eqref{eq:FF} for five (four) different points in the $q^2$ region $[-15,\,5]\,\gev^2$ for the form factors $f_{+,T}\, (f_0)$ for $B_s \to \eta_s$ and $B_s \to K$ modes, whereas for the $B_s \to D_s$ mode, we restrict ourselves till $q^2$ = 0 $\text{GeV}^2$ since the relative contribution from the higher-twist two-particle terms increases with increasing positive values of $q^2$ making the calculation of the form factors unstable, as discussed in Ref. \cite{Gubernari:2018wyi}. Subsequently we use the $z$-parametrization  for the extrapolation of form factors from negative $q^2$ to the semileptonic region using Eq.~\eqref{eq:zexpfit} with appropriate mass parameters. The central values and $\pm 1 \sigma$ CIs of the parameters $a_k^{(i)}$ are provided in Appendix~\ref{app:coeff} (Table \ref{tab:coefficients}), along with their correlations (Tables \ref{tab:corrBs2eta}, \ref{tab:corrBs2K} and \ref{tab:corrBs2Ds}). For the $B_s \to \eta_s$ channel, we now compare the form factor predictions with our previous result i.e., using $\lambda_{B_s}$ value from QCD sum rule estimate in Eq. \eqref{eq:val} in green bands with the extracted value in Eq.~\eqref{eq:lamval} in blue bands. All other parameters are chosen to be same as already discussed above.
 As expected, the form factor estimates using the $\lambda_{B_s}$ value from Eq. \eqref{eq:val} have larger uncertainties and are relatively higher as compared to the estimates using the obtained value of $\lambda_{B_s}$ in our case, though the results are highly consistent with each other. We also show the data points at $q^2$ = 0 $\text{GeV}^2$ from the HPQCD collaboration \cite{Parrott:2022rgu} in black, which agrees well with our estimates as expected.

\begin{figure*}[t!!!!!]
		\small
		\centering
\includegraphics[width=0.32\textwidth]{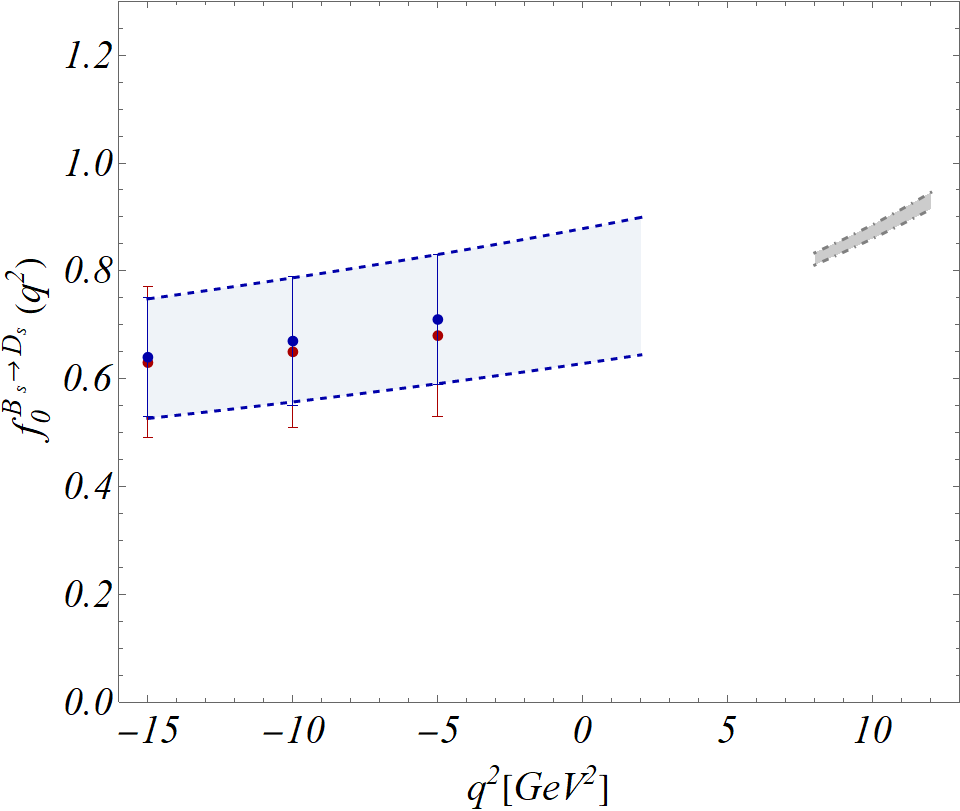}
  ~~
  \includegraphics[width=0.32\textwidth]{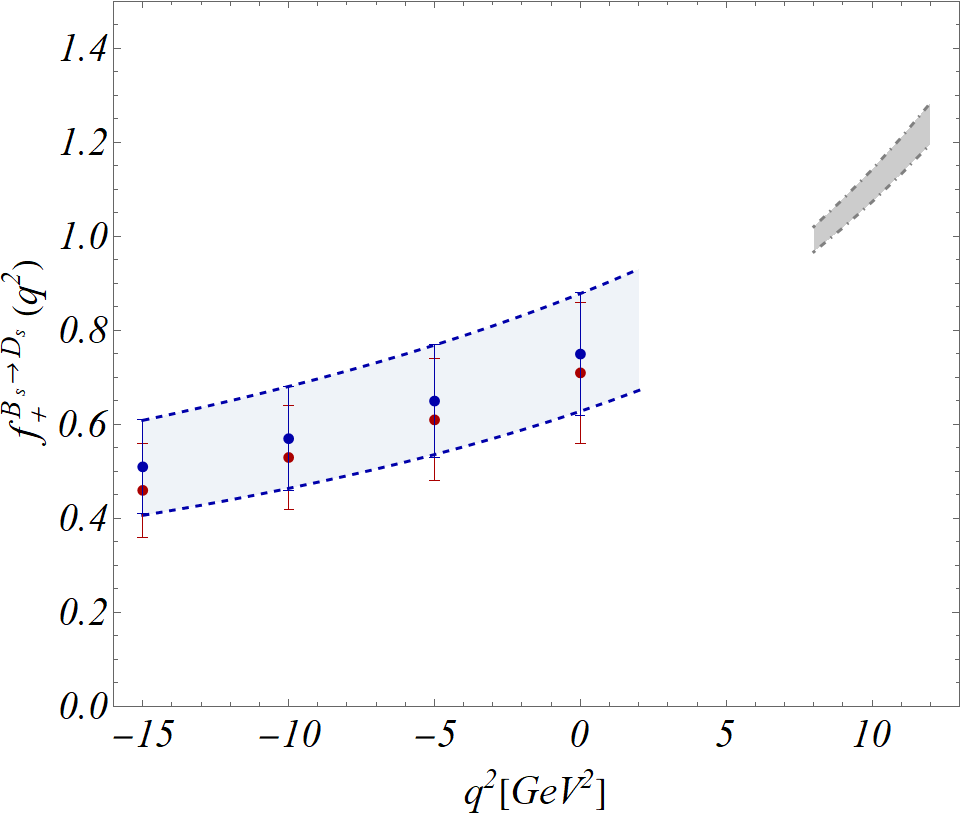}
		~~
		\includegraphics[width=0.32\textwidth]{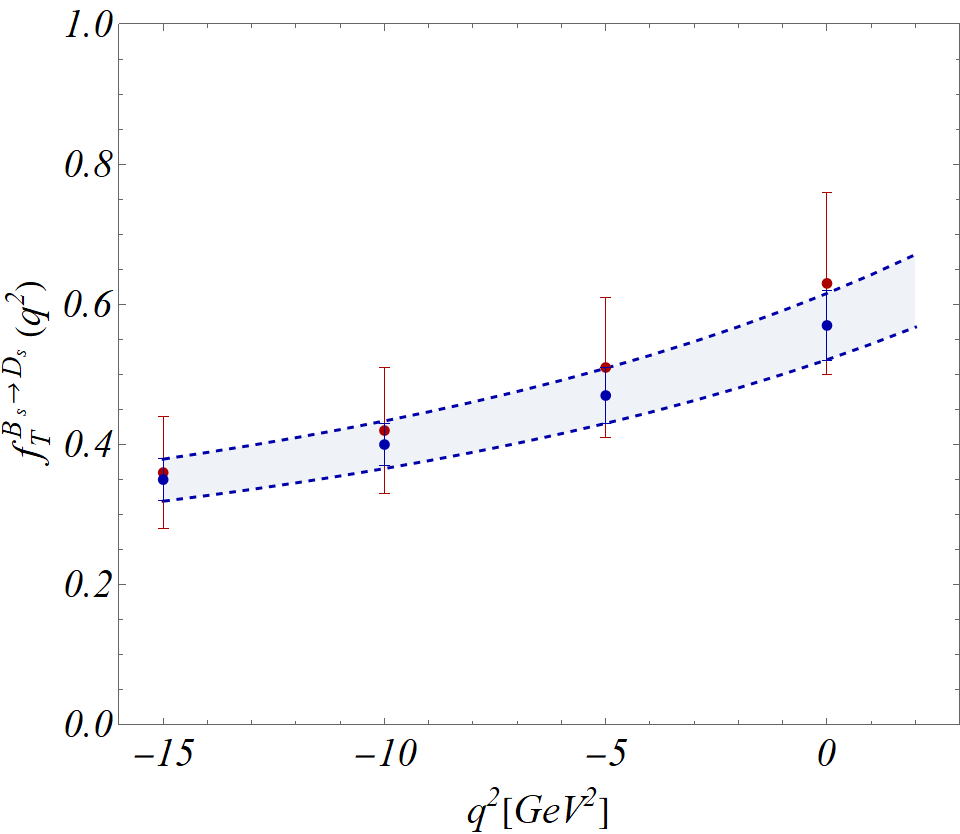}
		\\
\includegraphics[width=0.32\textwidth]{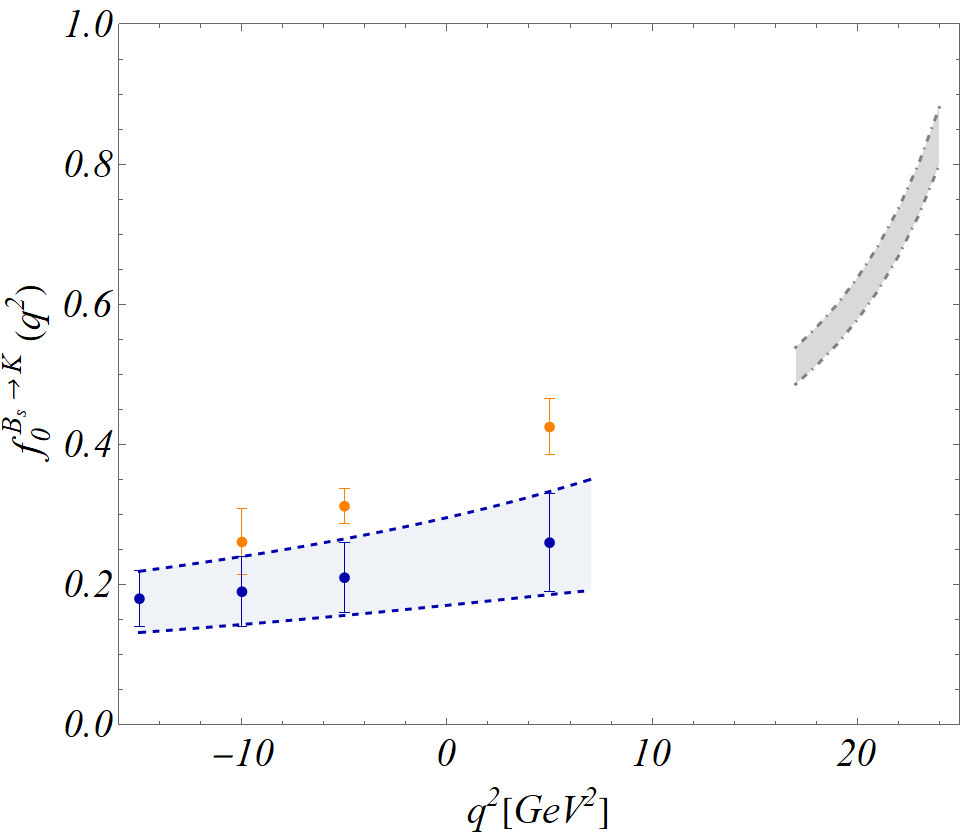}
  ~~
  \includegraphics[width=0.32\textwidth]{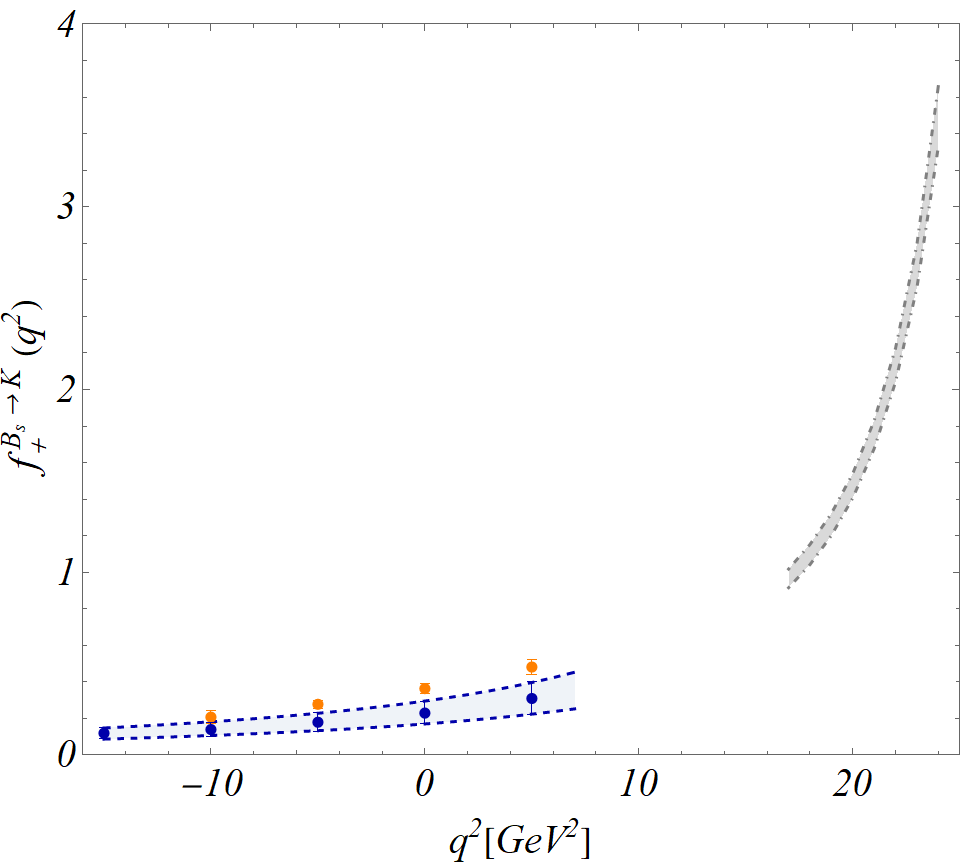}
		~~
		\includegraphics[width=0.32\textwidth]{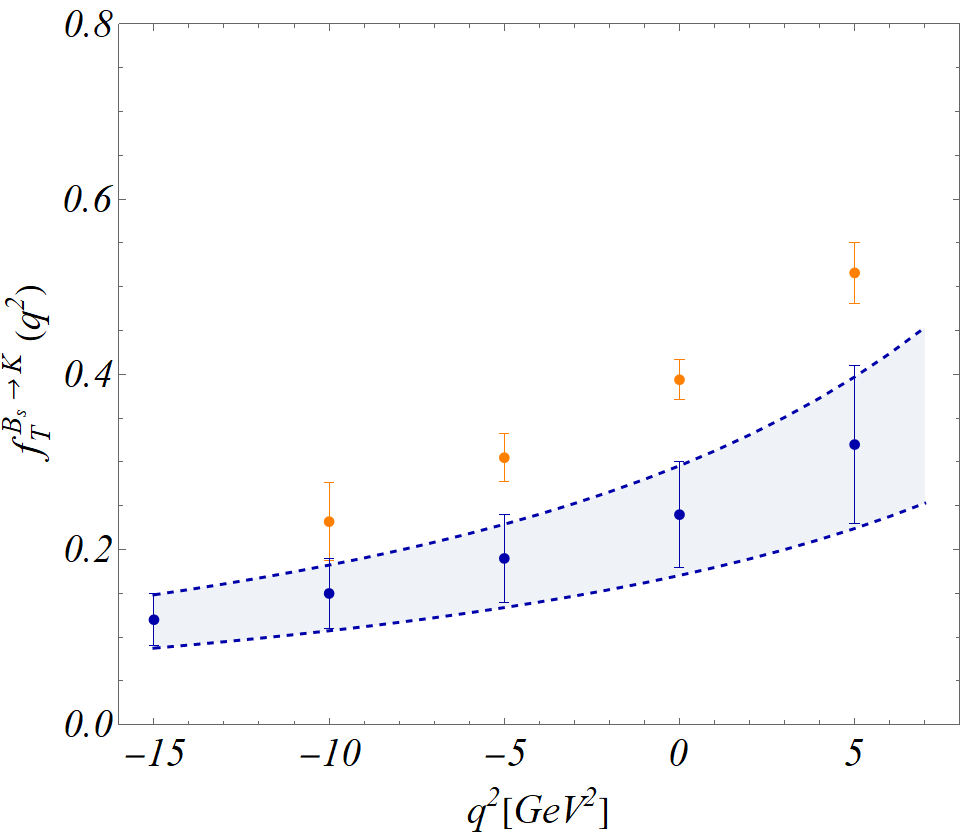}
		\\
  \includegraphics[width=0.3\textwidth]{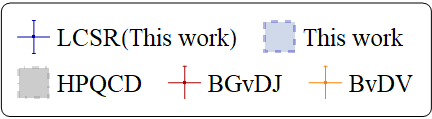}
		\caption{\small The variation of the form factors with momentum transfer $q^2$ for the modes $B_s \to D_s$ (upper row) and $B_s \to K$ (lower row). The blue band and error bars represent LCSR predictions using our extracted value of $\lambda_{B_s}$ (Eq.~\eqref{eq:lamval}). The red(orange) error bars correspond to the LCSR estimates from BGvDJ \cite{Bordone:2019guc}(BvDV \cite{Bolognani:2023mcf}). The grey bands correspond to Lattice QCD estimates from
HPQCD collaboration (see Ref. \cite{McLean:2019qcx} for $B_s \to D_s$ and Ref. \cite{Bouchard:2014ypa} for $B_s \to K$ decays.)}
		\label{fig:plots}
	\end{figure*}

For the $B_s \to D_s$ channel, we compare our results to those from Ref. \cite{Bordone:2019guc} [BGvDJ](red error bars), where $B$-meson LCDAs have been used ignoring the $SU(3)_F$ violation effect due to the presence of $s$-quark in the meson. Considering the values of the threshold and Borel parameters from [BGvDJ], we calculate the form factors with our extracted value of $\lBs$.  For the form factors $f_0$ and $f_+$, the values obtained in our case are slightly higher than [BGvDJ], whereas for the $f_T$ form factor, the pattern is opposite. For this channel also, the uncertainty for all the form factors are reduced and the results are highly consistent with each other. We have also shown Lattice QCD estimates from the HPQCD collaboration \cite{McLean:2019qcx} for $q^2$ $\gtrsim$ 8 $\text{GeV}^2$ as the lattice simulations for this mode are most precise in this region\footnote{The Lattice QCD estimates are available only for the form factors $f_+$ and $f_0$ for the $B_s \to (K,D_s)$ modes.}.

In the case of $B_s \to K$ channel, the predictions of form factors using $B_s$-meson LCDAs (with QCD sum rule estimate of $\lambda_{B_s}$ in Eq.~\eqref{eq:val}) bear with large uncertainty. However, the precise prediction for form factors are obtained using $K$-meson LCDAs in Ref. \cite{Bolognani:2023mcf} [BvDV] and thus we compare our results to those of [BvDV] (orange error bars). We have obtained the threshold parameters separately for the form factors $f_+$ and $f_T$ following the method as described above (for $B_s \to \eta_s$) resulting into $s_0$ = 0.58 $\pm$ 0.04 $\text{GeV}^2$ (0.63 $\pm$ 0.11  $\text{GeV}^2$) for form factors $f_+$($f_T$). We notice that the uncertainties of the form factors obtained using light-meson LCDAs is significantly reduced in comparison to our results, which is as per the expectation. We find that the central values obtained in [BvDV] are relatively higher than ours and are consistent only at $2\sigma$ CI for most of the region. These are two complementary methods to calculate the form factors and we have demonstrated that improvement in $B_s$-meson LCDA parameters will eventually lead to predictions with comparable uncertainty. We have also shown Lattice estimates from the HPQCD collaboration \cite{Bouchard:2014ypa} for $q^2$ $\gtrsim$ 16 $\text{GeV}^2$ region for the form factors $f_+$ and $f_0$. However, there is currently no estimate available for the form factor $f_T$, also from other Lattice collaborations~\cite{Flynn:2023nhi,FermilabLattice:2019ikx} in the same decay mode.

\section{Summary}
\label{sec:summary}

In this study, we utilize form factor results obtained from Lattice QCD concerning the $\eta_s$-meson. The $\eta_s$-meson, a fictitious $s\bar s$ bound state used to calibrate the $s$-quark mass in Lattice QCD analyses, is generally restricted in its application and has not been utilized elsewhere. Initially, we employ the QCD sum rule framework to compute the decay constant for the $\eta_s$-meson using two-point sum rules. Perturbative contributions are accounted for up to next-to-leading order, while condensate effects are considered up to dimension five operators. Notably, these contributions are largely proportional to the $s$-quark mass, except for the gluon condensate, which significantly influences the sum rule. Our prediction for the decay
constant is consistent with the Lattice QCD estimate and has a smaller degree of uncertainty.

With estimates of intrinsic parameters for QCD sum rules in hand, we proceed to compute the $B_s \to \eta_s$ form factors within the LCSR framework, using $B_s$-meson LCDAs. These analytical expressions for the form factors are then utilized to extract the inverse moment of the leading-twist $B_s$-meson LCDA, denoted as $\lBs$, leveraging recent Lattice QCD results obtained for these form factors at zero momentum transfer $(q^2 = 0~ \text{GeV}^2)$ from the HPQCD collaboration. Our analysis yields $\lambda_{B_s} = (480 \pm 92)\,\mev$ when employing the Exponential model for $B_s$-meson LCDA parametrization. Notably, this value is consistent with prior QCD sum rule estimate of $\lBs$, however exhibiting a $1.5$-fold reduction in uncertainty.

Using the indirectly determined value of $\lBs$, we proceed to compare the form factor estimates across various modes of $B_s$-meson decay, including $B_s \to D_s$ and $B_s \to K$. Our findings reveal consistency with previously obtained $B_s \to D_s$ form factors calculated using $B$-meson LCDAs, albeit with improved overall uncertainties in our analysis. However, comparisons with previous $B_s \to K$ form factor predictions, utilizing $K$-meson LCDAs, demonstrate significantly greater precision compared to estimates derived from $B_s$-meson LCDAs, despite the incorporation of the constrained value of $\lBs$ obtained in this work.

\subsection*{Acknowledgement:}
R.M. acknowledges SERB Grant SPG/2022/001238 for the support. I.R.'s work receives support from the IIT Gandhinagar Research and Development Grant IP/IITGN//PHY/RM/2223/11.  \\

\appendix

\section{$B_s$-meson light-cone distribution amplitudes}
\label{app:BDA}
	
 We  express the $B_s$-to-vacuum matrix element in the heavy quark effective theory limit in which the heavy $b$-quark field is substituted with the heavy quark effective theory field $h_v$, with $v^\mu=\left( q+k\right)^\mu/m_{B_s}$ representing the $B_s$-meson's four-velocity in its rest frame.
The expansion in terms of the different twists two-particle $B_s$-meson LCDAs is given as \cite{Braun:2017liq}
	\begin{align}
	\bra{0} \bar{s}^{\alpha}(x) h_{v}^{\beta}(0) \ket{\bar{B_s}(v)} =&
	-\frac{i f_{B_s} m_{B_s}}{4} \int^\infty_0 d\omega  \bigg\{
	(1 + \slashed{v}) \bigg[
	\phi_+(\omega)
	- g_+(\omega) \partial_\lambda \partial^\lambda + \frac12 \left(\overline{\phi}(\omega)- \overline{g}(\omega) \right. \nn \\ & \left. \hspace{5cm}
	 \partial_\lambda \partial^\lambda\right) 
	\gamma^\rho \partial_\rho
	\bigg] \gamma_5
	\bigg\}^{\beta\alpha} e^{-i r \cdot x}
	\Bigg|_{r=\omega v}
	\,.
	\label{eq:BLCDAs2pt}
	\end{align}
	The derivatives $\partial_\mu \equiv \partial/\partial r^\mu$ are understood to act on the hard-scattering kernel.

	The Exponential model for $B_s$-meson LCDAs is given as 
	\cite{Grozin:1996pq,Braun:2017liq}
	\begin{align}
	\phi_+(\omega) &= \frac{\omega}{\lambda_{B_s}^2}e^{-\omega/\lambda_{B_s}}, \\
	\phi_-(\omega)
	&= \frac{1}{\lambda_{B_s}}e^{-\omega/\lambda_{B_s}} - \frac{\lambda_E^2 - \lambda_H^2}{18 \lambda_{B_s}^5} \left(2 \lambda_{B_s}^2 - 4 \omega \lambda_{B_s} + \omega^2 \right)e^{-\omega/\lambda_{B_s}}\,,\\
	g_+(\omega)&=
	-\frac{\lambda_E^2}{6\lambda_{B_s}^2}
	\biggl\{(\omega -2 \lambda_{B_s})  \text{Ei}\left(-\frac{\omega}{\lambda_{B_s}}\right) +  
	(\omega +2\lambda_{B_s}) e^{-\omega/\lambda_{B_s}}
	\times \left(\ln \frac{\omega}{\lambda_{B_s}}+\gamma_E\right)-2 \omega e^{-\omega/\lambda_{B_s}}\biggr\} \nn \\
 &+ \frac{e^{-\omega/\lambda_{B_s}}}{2{ \lambda_{B_s}}}\omega^2\biggl\{1 - \frac{1}{36\lambda_{B_s}^2}(\lambda_E^2- \lambda_H^2)\biggr\}, \label{gplus-model1} \\
 g_-^{WW}(\omega)
        & =  \frac{1}{4} \int_0^{\omega} \mathrm{d}\eta_2 \, \int_0^{\eta_2} \mathrm{d}\eta_1 \, \left[ \phi_+ (\eta_1) - \phi_-^{WW} (\eta_1) \right]
        - \frac{1}{2} \int_0^{\omega} \mathrm{d}\eta_1 \, \left(\eta_1 - \frac{3 }{2}\lBs\right) \phi_-^{WW} (\eta_1)  \nn \\
        & =  \frac{3 \omega}{4}\,  e^{-\omega/\lambda_{B_s}} \,.
	\label{eq:gpmWW}
	\end{align}
	where $\phi_-^{WW}(\eta_1)
	= e^{-\eta_1/\lambda_{B_s}}/ \lambda_{B_s}$ and $\text{Ei}(x)$ is the exponential integral. The approximate expression for $g_-$ is derived in the Wandzura-Wilczek limit, as mentioned in Ref.~\cite{Gubernari:2018wyi} where the three-component amplitudes are neglected (see also Ref. \cite{Kawamura:2001jm})\footnote{The LCDA $g_-$ being of twist five is suppressed as compared to the leading twist LCDAs and, thus its contribution in the form factors is numerically insignificant.}.

 	\section{Coefficients of the LCSR formula} \label{app:C_coeff}
	Here, we provide a listing of all the coefficients of the two-particle LCDAs that are involved in Eq. \eqref{eq:FF}. With the introduction of following notation
 \begin{align}
     f_\pm= \frac{q^2 - m_{B_s}^2 + m_P^2}{q^2} f_+ + \frac{m_{B_s}^2 - m_P^2}{q^2} f_0\,,
\label{eq:fpm}
 \end{align}
we obtain, for $f_+^{B_s \to P}$: 
	
	\begin{equation}
	\begin{aligned}
	C^{(f_+^{B_s \to P},\phi_+)}_1     & = -\bar{\sigma}\,,  \\
	C^{(f_+^{B_s \to P},\bar{\phi})}_2 & = -m_{B_s}\bar{\sigma}^2\,, \\
	C^{(f_+^{{B_s} \to P},g_+)}_2        & = -4\bar{\sigma}, \\
	C^{(f_+^{{B_s} \to P},g_+)}_3        & = 8 m_{q_1}^2 \bar{\sigma}\,, \\
	C^{(f_+^{{B_s} \to P},\bar{g})}_3    & = -8 m_{B_s} \bar{\sigma}^2\,, \\
	C^{(f_+^{{B_s} \to P},\bar{g})}_4    & = 24 m_{q_1}^2 m_{B_s}\bar{\sigma}^2\,.
	\end{aligned}
	\end{equation}
	For $f_\pm^{{B_s} \to P}$:
	\begin{equation}
	\begin{aligned}
	C^{(f_\pm^{{B_s} \to P},\phi_+)}_1     & = 2\sigma-1\,,                              \\
	C^{(f_\pm^{{B_s} \to P},\bar{\phi})}_2 & = 2 m_{B_s}\sigma\bar{\sigma}-m_{q_1}\,,            \\
	C^{(f_\pm^{{B_s} \to P},g_+)}_2        & = 4(2\sigma-1)\,,                           \\
	C^{(f_\pm^{{B_s} \to P},g_+)}_3        & = -8 m_{q_1}^2 (2\sigma-1)\,,                     \\
	C^{(f_\pm^{{B_s} \to P},\bar{g})}_3    & = 16 m_{B_s} \sigma\bar{\sigma}\,,             \\
	C^{(f_\pm^{{B_s} \to P},\bar{g})}_4    & = 24 m_{q_1}^2(m_{q_1}-2 m_{B_s} \sigma\bar{\sigma})\,.
	\end{aligned}
	\end{equation}
	For $f_T^{{B_s} \to P}$:
	\begin{equation}
	\begin{aligned}
	C^{(f_T^{{B_s} \to P},\bar{\phi})}_1 &= \frac{1}{m_{B_s}}, \\
  C^{(f_T^{{B_s} \to P},\bar{\phi})}_2 &= \frac{-(m_{B_s}^2 \bar{\sigma}^2-m_{q_1}^2+2q^2\sigma-q^2)}{m_{B_s}},& \\
	 C^{(f_T^{{B_s} \to P},\bar{g})}_2 &= \frac{8}{m_{B_s}}, \\
  C^{(f_T^{{B_s} \to P},\bar{g})}_3 & = \frac{-8(m_{B_s}^2 \bar{\sigma}^2+2 m_{q_1}^2+2q^2\sigma-q^2)}{m_{B_s}},\\
	C^{(f_T^{{B_s} \to P},\bar{g})}_4&= \frac{24 m_{q_1}^2(m_{B_s}^2 \bar{\sigma}^2-m_{q_1}^2+2q^2\sigma-q^2)}{m_{B_s}}.
	\end{aligned}
	\end{equation}
	
	The normalization factors (in Eq. \eqref{eq:FF}) are,
	\begin{equation}
 \label{eq:Kfac}
	K^{(f_+^{{B_s} \to P})}=K^{(f_\pm^{{B_s} \to P})}=f_P, ~~
	K^{(f_T^{{B_s} \to P})}=\frac{f_P(m^2_{B_s}-m^2_P-q^2)}{m_{B_s}(m_{B_s}+m_P)}\,.
	\end{equation}
In the above expressions, the quark mass $m_{q_1}$ enters for different transitions with $q_1=s,\,d$ and $c$ for the modes $B_s \to \eta_s,\,K\,\text{and}\,D_s$, respectively.

\section{Fitted coefficients and the correlations}
\label{app:coeff}

In Table \ref{tab:coefficients}, we present the results for the fitted expansion coefficients, denoted as $a_k^{(i)}$ in Eq.~\eqref{eq:zexpfit}, for the various modes considered in this study. The respective correlations among the fitted coefficients are provided in Tables \ref{tab:corrBs2eta}, \ref{tab:corrBs2K}, and \ref{tab:corrBs2Ds}.

\begin{table*} [h!!!!!]
		\small
		\centering
		\renewcommand*{\arraystretch}{1.3}
		\begin{tabular}{|c||c|c|c|}
			\hline
			& $B_s \to \eta_s$ & $B_s \to K$ & $B_s \to D_s$ \\
			\hline
			$a^+_0$ & 0.31 $\pm$ 0.11 & 0.23 $\pm$ 0.06 & 0.75 $\pm$ 0.13 \\
			$a^+_1$ & -0.71 $\pm$ 0.23 & -0.59 $\pm$ 0.21 & -0.86 $\pm$ 0.49 \\
			$a^+_2$ & -0.31 $\pm$ 0.49 & -0.13 $\pm$ 0.35 & -0.69 $\pm$ 1.86 \\
			$a^0_1$ & 0.56 $\pm$ 0.19 & 0.39 $\pm$ 0.04 & 2.00 $\pm$ 0.48 \\
			$a^0_2$ & -1.03 $\pm$ 0.22 & -0.74 $\pm$ 0.05 & -1.82 $\pm$ 0.56 \\
			$a^T_0$ & 0.32 $\pm$ 0.07 & 0.24 $\pm$ 0.06 & 0.57 $\pm$ 0.05 \\
			$a^T_1$ & -0.79 $\pm$ 0.18 & -0.62 $\pm$ 0.21 & -2.03 $\pm$ 0.47  \\
			$a^T_2$ & 0.002 $\pm$ 0.48 & -0.06 $\pm$ 0.38 & 9.51 $\pm$ 2.29 \\
			\hline	\end{tabular}
		\caption{\small $z$-expansion coefficients for the various channels considered in this work.}
		\label{tab:coefficients}
	\end{table*}

\begin{table*} [h!!!!!]
		\small
		\centering
		\renewcommand*{\arraystretch}{1.3}
		\begin{tabular}{|c||cccccccc|}
			\hline
			& $a^+_0$ & $a^+_1$ & $a^+_2$ & $a^0_1$ & $a^0_2$ & $a^T_0$ & $a^T_1$ & $a^T_2$ \\
\hline
$a^+_0$ & $1.$  &  $-0.89$  &  $-0.117$  &  $0.885$  &  $-0.497$  &  $0.1$  &  $-0.061$  &  $-0.005$  \\
\hline
$a^+_1$ & $-0.89$  &  $1.$  &  $-0.341$  &  $-0.593$  &  $0.196$  &  $-0.067$  &  $-0.024$  &  $0.123$  \\
\hline
$a^+_2$ & $-0.117$  &  $-0.341$  &  $1.$  &  $-0.544$  &  $0.691$  &  $-0.034$  &  $0.155$  &  $-0.262$  \\
\hline
$a^0_1$ & $0.885$  &  $-0.593$  &  $-0.544$  &  $1.$  &  $-0.793$  &  $0.111$  &  $-0.122$  &  $0.104$  \\
\hline
$a^0_2$  & $-0.497$  &  $0.196$  &  $0.691$  &  $-0.793$  &  $1.$  &  $-0.064$  &  $0.097$  &  $-0.143$  \\
\hline
$a^T_0$ & $0.1$  &  $-0.067$  &  $-0.034$  &  $0.111$  &  $-0.064$  &  $1.$  &  $-0.83$  &  $0.208$  \\
\hline
$a^T_1$ & $-0.061$  &  $-0.024$  &  $0.155$  &  $-0.122$  &  $0.097$  &  $-0.83$  &  $1.$  &  $-0.693$  \\
\hline
$a^T_2$ & $-0.005$  &  $0.123$  &  $-0.262$  &  $0.104$  &  $-0.143$  &  $0.208$  &  $-0.693$  &  $1.$  \\
\hline
\end{tabular}
\caption{\small Correlation among the $z$-expansion coefficients for $B_s \to \eta_s$ mode.} 
		\label{tab:corrBs2eta}
	\end{table*}

\begin{table*} [h!!!!!]
		\small
		\centering
		\renewcommand*{\arraystretch}{1.3}
		\begin{tabular}{|c||cccccccc|}
			\hline
			& $a^+_0$ & $a^+_1$ & $a^+_2$ & $a^0_1$ & $a^0_2$ & $a^T_0$ & $a^T_1$ & $a^T_2$ \\
\hline
$a^+_0$ &
 $1.$  &  $-0.988$  &  $0.88$  &  $0.73$  &  $0.243$  &  $0.011$  &  $-0.013$  &  $0.015$  \\
\hline
$a^+_1$ & $-0.988$  &  $1.$  &  $-0.943$  &  $-0.623$  &  $-0.366$  &  $-0.005$  &  $0.005$  &  $-0.004$  \\
\hline
$a^+_2$ & $0.88$  &  $-0.943$  &  $1.$  &  $0.342$  &  $0.606$  &  $-0.011$  &  $0.015$  &  $-0.022$  \\
\hline
$a^0_1$ & $0.73$  &  $-0.623$  &  $0.342$  &  $1.$  &  $-0.468$  &  $0.027$  &  $-0.039$  &  $0.057$  \\
\hline
$a^0_2$ & $0.243$  &  $-0.366$  &  $0.606$  &  $-0.468$  &  $1.$  &  $-0.032$  &  $0.045$  &  $-0.066$  \\
\hline
$a^T_0$ & $0.011$  &  $-0.005$  &  $-0.011$  &  $0.027$  &  $-0.032$  &  $1.$  &  $-0.973$  &  $0.793$  \\
\hline
$a^T_1$ & $-0.013$  &  $0.005$  &  $0.015$  &  $-0.039$  &  $0.045$  &  $-0.973$  &  $1.$  &  $-0.912$  \\
\hline
$a^T_2$ & $0.015$  &  $-0.004$  &  $-0.022$  &  $0.057$  &  $-0.066$  &  $0.793$  &  $-0.912$  &  $1.$  \\
\hline
\end{tabular}
\caption{\small Correlation among the $z$-expansion coefficients for $B_s \to K$ mode.} 
		\label{tab:corrBs2K}
	\end{table*}

\begin{table*} [h!!!!!]
		\small
		\centering
		\renewcommand*{\arraystretch}{1.3}
		\begin{tabular}{|c||cccccccc|}
			\hline
			& $a^+_0$ & $a^+_1$ & $a^+_2$ & $a^0_1$ & $a^0_2$ & $a^T_0$ & $a^T_1$ & $a^T_2$ \\
\hline
$a^+_0$ &
 $1.$  &  $0.556$  &  $-0.673$  &  $0.855$  &  $0.153$  &  $0.256$  &  $0.698$  &  $-0.539$  \\
\hline
$a^+_1$ & $0.556$  &  $1.$  &  $-0.883$  &  $0.89$  &  $0.009$  &  $-0.533$  &  $0.869$  &  $-0.519$  \\
\hline
$a^+_2$ & $-0.673$  &  $-0.883$  &  $1.$  &  $-0.887$  &  $0.341$  &  $0.246$  &  $-0.932$  &  $0.805$  \\
\hline
$a^0_1$ & $0.855$  &  $0.89$  &  $-0.887$  &  $1.$  &  $0.044$  &  $-0.128$  &  $0.846$  &  $-0.561$  \\
\hline
$a^0_2$ & $0.153$  &  $0.009$  &  $0.341$  &  $0.044$  &  $1.$  &  $-0.132$  &  $-0.113$  &  $0.495$  \\
\hline
$a^T_0$ & $0.256$  &  $-0.533$  &  $0.246$  &  $-0.128$  &  $-0.132$  &  $1.$  &  $-0.383$  &  $-0.002$  \\
\hline
$a^T_1$ & $0.698$  &  $0.869$  &  $-0.932$  &  $0.846$  &  $-0.113$  &  $-0.383$  &  $1.$  &  $-0.795$  \\
\hline
$a^T_2$ & $-0.539$  &  $-0.519$  &  $0.805$  &  $-0.561$  &  $0.495$  &  $-0.002$  &  $-0.795$  &  $1.$  \\
\hline
\end{tabular}
\caption{\small Correlation among the $z$-expansion coefficients for $B_s \to D_s$ mode.} 
		\label{tab:corrBs2Ds}
	\end{table*}

\bibliographystyle{bibstyle}
\bibliography{example.bib}

\end{document}